\documentclass[%
 aip,
 amsmath,amssymb,
 reprint,%
]{revtex4-1}

\usepackage[pdftex]{graphicx}
\usepackage{dcolumn}
\usepackage{bm}

\usepackage{subcaption} 

\usepackage[utf8]{inputenc}
\usepackage[T1]{fontenc}
\usepackage{mathptmx}
\usepackage{etoolbox}
\usepackage{float}
\usepackage{hyperref}
\usepackage[ruled]{algorithm2e}
\RestyleAlgo{ruled}

\SetCommentSty{mycommfont}

\newcommand{\nosemic}{\renewcommand{\@endalgocfline}{\relax}}
\newcommand{\dosemic}{\renewcommand{\@endalgocfline}{\algocf@endline}}

\makeatletter
\def\@email#1#2{%
 \endgroup
 \patchcmd{\titleblock@produce}
  {\frontmatter@RRAPformat}
  {\frontmatter@RRAPformat{\produce@RRAP{*#1\href{mailto:#2}{#2}}}\frontmatter@RRAPformat}
  {}{}
}%
\makeatother
\begin{document}

\preprint{AIP/123-QED}

\title {Prior Global Search Stability on Finite Graphs with Uncertainty. May Greedy Search Win?}

\author{Andrey Ananev}
 \altaffiliation{ananev@phystech.edu}
\affiliation {Moscow Institute of Physics and Technology, Institutskiy Pereulok 9, Dolgoprudnyy, 141701, Russia}

\author{Aleksey Khlyupin}%
 \altaffiliation{khlyupin@phystech.edu}
\affiliation{Moscow Institute of Physics and Technology, Institutskiy Pereulok 9, Dolgoprudnyy, 141701, Russia}

\begin{abstract}

This research paper addresses the stability of search algorithms in complex networks when dealing with incomplete information or uncertainty. We propose a theoretical model to investigate whether a global search algorithm with incomplete prior information can be outperformed by a stochastic greedy search on average. The model incorporates random variables to perturb edge weights in the graph, thus capturing the uncertainty of available information. Our findings indicate that some graphs and uncertainty model parameters exist where the global search algorithm fails under uncertainty conditions, while the random greedy search performs better. We derive a critical curve that separates stable from unstable graphs for global search with incomplete information. Interestingly, the critical curve's behavior changes from monotonic to bell-shaped depending on the uncertainty parameters. We test our proposed model through numerical simulations on various synthetic and real-world graphs with different structures. Our results offer insights into the design and optimization of search algorithms for network-based applications, such as communication networks, social networks, and biological networks. We also discuss the study of memory and associative learning in miniature insects, highlighting the potential of efficient search and walking strategies for small robots or devices that operate in a limited area in space.

\end{abstract}

\maketitle

\section{\label{sec:introduction}Introduction}
Imagine an autonomous vehicle navigating the bustling streets of a dynamic urban landscape. To reach its destination efficiently, it must not only adapt to real-time traffic but also anticipate the unexpected—road closures, accidents, or sudden route changes. Autonomous systems navigating unpredictable urban terrains, adaptive networks routing information through an ever-changing internet, or biological agents optimizing search strategies in spatially constrained environment, in each scenario, the challenge is clear — efficiently charting a course through uncertainty. The success of such intelligent agents hinges on their ability to make decisions in ever-evolving environments. These scenarios epitomize the profound challenge of navigating complex networks amidst incomplete information and uncertainty, a challenge that reverberates through various facets of our modern world.

Complex networks are the connective tissue of our interconnected world, weaving through diverse fields, from communication systems [\onlinecite{boguna2009navigability}] to social interactions [\onlinecite{eckmann2004entropy}, \onlinecite{palchykov2012sex} \onlinecite{zhao2011social}] and biological processes [\onlinecite{viswanathan2000levy}, \onlinecite{weiss2014use}]. A fundamental challenge in the study of these networks is the development of search algorithms that can efficiently navigate through them, especially in the presence of incomplete information or uncertainty. These algorithms serve as the guiding intelligence, enabling efficient traversal of complex networks, all while grappling with the inherent ambiguities and unpredictabilities of real-world data.

In the realm of graph theory, algorithms for finding the optimal path have long been known. Dijkstra's algorithm, introduced by Edsger W. Dijkstra in 1956 [\onlinecite{dijkstra2022note}], efficiently determines the shortest path in weighted graphs and has found applications in various domains, from transportation networks to computer and social networks [\onlinecite{wang2011application}, \onlinecite{fan2010improvement}]. Other algorithms like the A* algorithm [\onlinecite{hart1968formal}] combine Dijkstra's principles with heuristics to find optimal paths efficiently, particularly in artificial intelligence and video game pathfinding [\onlinecite{cui2011based}, \onlinecite{hutchison2009algorithmics}, \onlinecite{zeng2009finding}]. While classical pathfinding algorithms have made significant progress, they necessitate prior knowledge of all information concerning the graph in which the pathfinding is conducted.

This research paper delves into the intricate dynamics of search algorithms operating within complex networks, with a particular focus on their stability and performance under conditions of uncertainty. In essence, we seek answers to a critical question: Can a global search algorithm, armed with incomplete prior information, consistently outperform a stochastic greedy search algorithm concerning the identification of the shortest route on average within these uncertain conditions? Answering this question is far from trivial, as it involves a nuanced analysis of the interplay between search strategies, network topology, and the inherent uncertainty of the information available.

To address this challenge, we present a robust theoretical model designed to explore the intricate relationship between search algorithms and network uncertainty. This model introduces a key element of uncertainty by perturbing the edge weights in network graphs using random variables. These variables serve as proxies for the uncertain nature of available information in the real world. Our findings reveal that, indeed, certain scenarios exist where global search algorithms falter under conditions of uncertainty, while stochastic greedy searches prove to be more effective. These results fundamentally alter our understanding of search dynamics within complex networks and illuminate the stability of search algorithms in the face of incomplete information.

\begin{figure*}[t]
\includegraphics[width=1\textwidth]{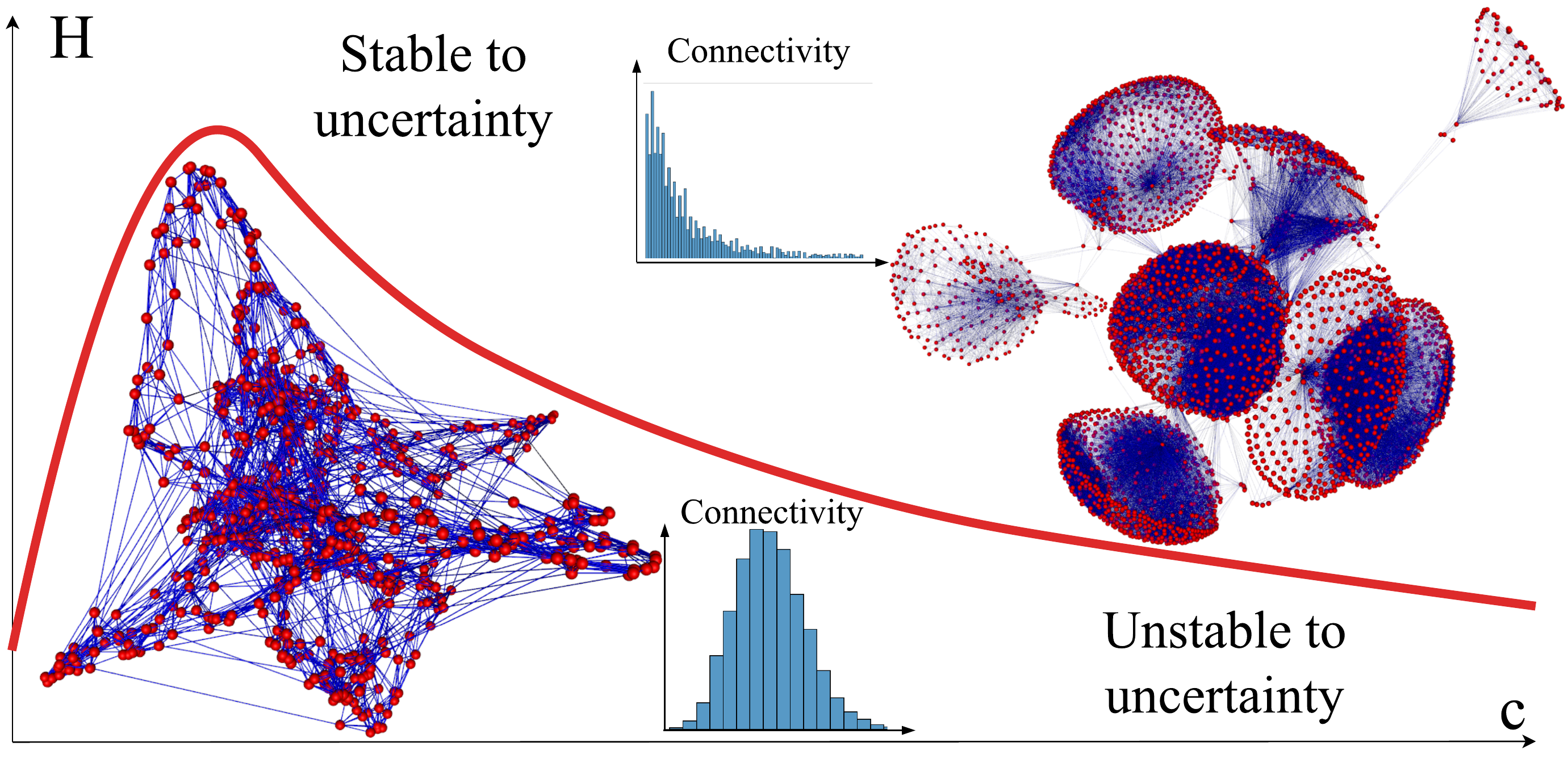}
\caption{\label{fig:graphical_abstract} The figure shows the main purpose of the article:
show the influence of the graph structure (with diameter $2H$ and connectivity $c$) on the resistance to different types of search algorithms on the graph (greedy or global) under uncertainty.}
\end{figure*}

A pivotal outcome of our investigation is the identification of a critical curve, a distinctive feature that distinguishes stable network configurations from unstable ones when employing global search with incomplete information. Intriguingly, this critical curve exhibits diverse behaviors, ranging from monotonic to bell-shaped, contingent on the specific parameters of uncertainty within the network. This discovery underscores the intricate and nuanced responses of networked systems to uncertainty, revealing both challenges and opportunities.

To substantiate our theoretical revelations, we embark on extensive numerical simulations encompassing a variety of network models. Our investigations traverse synthetic networks, including random regular [\onlinecite{wormald1999models}], n-dimensional uniform lattices [\onlinecite{eppstein2005lattice}], Small-world networks [\onlinecite{newman1999scaling},\onlinecite{watts1998collective}], Erd\H{o}s-R'enyi graphs [\onlinecite{erdHos1960evolution}], and Scale-free Barab'asi-Albert networks [\onlinecite{albert2002statistical}]. Additionally, we explore real-world networks, delving into a social network extracted from anonymized Facebook data [\onlinecite{facebook_dataset}] and a graph representing the vascular system of the mouse brain [\onlinecite{schneider2012tissue}]. Through meticulous simulations, we validate our theoretical framework, achieving impressive quantitative accord across numerous network models, with intriguing exceptions within networks characterized by heavy-tailed connectivity distributions and complex community structures [\onlinecite{dorogovtsev2002pseudofractal}, \onlinecite{krapivsky2000connectivity}]. The remarkable robustness of global search observed in such networks, despite theoretical predictions suggesting otherwise, underscores the need for advanced model development tailored to these specific scenarios.

Beyond the theoretical realm, this study resonates with practical implications across network science, graph theory, and optimization. The critical curve's behavior and its correlation with different network models and uncertainty parameters offer invaluable insights into the stability of search algorithms operating in complex networks under conditions of incomplete information or uncertainty. Moreover, this analysis holds promise for informing the design and optimization of network architectures across diverse domains, including communication networks [\onlinecite{farahani2013review}, \onlinecite{chandrasekharan2016designing}], social networks [\onlinecite{chen2010scalable}, \onlinecite{borgatti2018analyzing}, \onlinecite{chen2009efficient}], and biological networks [\onlinecite{wang2001bio}]. 

Our research delves into the intricate interplay between network search algorithms and uncertainty, shedding light on their stability and performance. By bridging theory and simulation across diverse network models, we uncover fundamental insights into networked systems, with implications ranging from optimizing search algorithms to illuminating the behavior of miniature organisms [\onlinecite{PRXLife.1.023001}, \onlinecite{polilov2015small}, \onlinecite{polilov2023extremely}] and guiding the design of advanced robotic systems.

\section{\label{sec:search_models_under_uncertainty}Search Models under Uncertainty in Networks}

In this section, two graph search models for graphs with incomplete a priori information about edge weights will be explored. The analysis will be restricted to finite graphs $G = (V,E,W)$, where $V$ represents the set of $N$ vertices and the set of edges and weights given on these edges will be denoted as $E$ and $W$, respectively. The search space's primary characteristic is that the actual edge weights might not match their a priori information and are perturbed by random variables $\xi_e$. In the probabilistic model $\mathbb{P}_\xi$, it is assumed that $\xi_e = 0$ with a high probability and does not alter the initial edge weights. However, with a low probability $p$, the random variable $\xi_e$ can take on large values $u$. Thus the actual weight of an edge in a certain implementation can be determined as follows:
\begin{equation}
\label{weight_xi}
    w_e = w^p_e + \xi_e
\end{equation}
where $w^p_e$ is priori weight of edge.
Two search models will be described below. The first model, referred to as the Dijkstra searcher or D-search, has complete a priori information about the edge weights and builds the optimal path between the source vertex $s$ and target vertex $t$ using a global search algorithm. The D-search then follows this path. The second search model, referred to as the Greedy searcher or G-search, implements a greedy search strategy from the same source vertex $s$. At each moment, the G-search only has access to the current edge weights incident with its current position and chooses the edge with the minimum weight to determine its next move.

The D-searcher can encounter several large random variables $\xi_e$ along the path built from a priori information. Each searcher walks a different path $Path(s,t)$ with a different weight $W(s,t)$, which defined as:
\begin{equation}
\label{path_acc}
    W_{s,t} = \sum_{e \in Path(s,t)} w_e 
\end{equation}

Our goal is to study the theoretical average of the route weight $\mathbb{E}[W_{s,t}]$ for these two search strategies and investigate whether the G-searcher can outperform the D-searcher. Exploring the graphs and parameters of uncertainty can lead to a better outcome for the G-searcher.


\begin{figure*}
\includegraphics[width=1\textwidth]{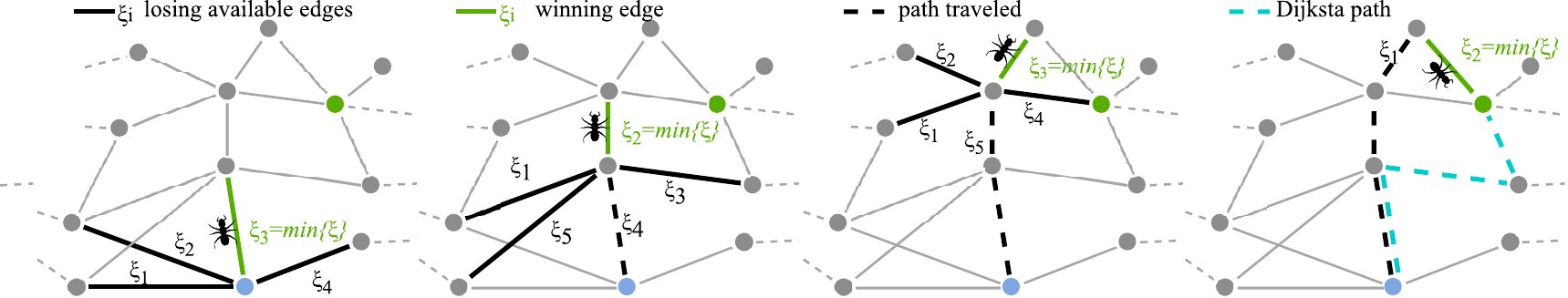}
\caption{\label{fig:greedy_ex} An example of a search path from source (blue vertex) to target (green vertex) using a G-searcher model. The searcher is represented in the picture as an ant. The green color of the edge indicates the route selected by the searcher in the current vertex, the remaining edges, whose weight is greater, are marked as black. The dotted line represents the founded path. }
\end{figure*}

\subsection{\label{sec:dsearcher}D-searcher}

As previously mentioned, the D-searcher does not take into account the uncertainty of weight and finds the shortest path using only a priori information (figure \ref{fig:greedy_ex}e). Let us denote the number of edges in the shortest path between vertices $s$ and $t$ as $L$. Then, the path weight $W^D_{s,t}$ can be calculated using forward substitution, from equation (\ref{weight_xi}) to (\ref{path_acc}).

\begin{equation}
    W^D_{s,t} = \sum_{e} (1 + \xi_e) = L + \sum_{e} \xi_e
    \label{dijkstra_wst}
\end{equation}
For ease of analysis, it is assumed that the a priori weight of the edges is 1.($w^p_e = 1$).
Here, the index $D$ refers to the weight of D-searcher path, and for simplicity, the sum over the edges of the path is denoted as $\sum_{e \in Path(s,t)} = \sum_{e}$. The expectation of the path weight can be expressed as:
\begin{equation}
    \mathbb{W}^{D}_{s,t} = \mathbb{E}[W^D_{s,t}] = L + \sum_{e}(\mathbb{E}_{\xi}) = L(1 + \mathbb{E}_{\xi})
\end{equation}
where $\mathbb{E}_{\xi}$ is the expectation of the random variable $\xi_e$.

\subsection{\label{sec:gsearcher}G-searcher}

Random walks, characterized by stochastic movement through a given space, are a fundamental concept that has traversed scientific history. Originating from Brown's observations in 1828, their formalization by Pearson in 1905 marked the beginning of a journey spanning diverse fields [\onlinecite{brown1828xxvii}, \onlinecite{pearson1905problem}]. These seemingly random processes have left an indelible mark in various disciplines, describing how our brains work when making decisions [\onlinecite{usher2001time}, \onlinecite{gold2007neural}]; polymer chains [\onlinecite{fisher1966shape}, \onlinecite{isichenko1992percolation}] and physics of surface phenomena [\onlinecite{khlyupin2021branching}]; descriptions of financial markets [\onlinecite{mantegna1999introduction}, \onlinecite{campbell2012econometrics}]; scientific citation network analysis [\onlinecite{rosvall2008maps},\onlinecite{jia2017quantifying}], construction of ranking systems [\onlinecite{fouss2007random},\onlinecite{gleich2015pagerank}]; simulation of fluid propagation in pore network models [\onlinecite{bijeljic2006pore}] or estimation of the rate of spread of disease in an organism or population [\onlinecite{christley2005infection},\onlinecite{navlakha2010power}]. Our focus here extends beyond traditional applications of random walk models, delving into the profound question of search algorithm efficiency within complex networks with incomplete information or uncertainty.

The G-searcher model does not require knowledge of the entire graph structure, only local information about its current position. This search model follows simple principles:
\begin{enumerate}
    \item Start at the current vertex ($v$);
    \item \label{get_weight} Get the weights of all incident edges $\{w_{v,v_1}, w_{v,v_2},...\}$;
    \item Find the edge $e_{v,l}$ with the minimum weight, and if there are several such edges, choose a random one;
    \item Move to vertex $l$, and add it to the path;
    \item Repeat from step \ref{get_weight} until the target vertex is found.
\end{enumerate}
In this search model, it is crucial that the random variable $\xi_e$ is updated on each iteration, even if the searcher has visited the vertex before (figure \ref{fig:greedy_ex}a, \ref{fig:greedy_ex}b, \ref{fig:greedy_ex}c, \ref{fig:greedy_ex}d). Based on the third step, the weight of the edge in the path of the G-searcher is denoted as the random variable $\varepsilon$.
\begin{equation}
    \varepsilon = \min \{ \xi_1, \xi_2, ... \}
\end{equation}
The expectation of the path weight can then be calculated as:
\begin{equation}
    \mathbb{W}^{G}_{s,t} = \sum_n n (\mathbb{E}_\varepsilon + 1) P_{st}(n|L)
    \label{w_greedy}
\end{equation}
Here, $P_{st}(n|L)$ is the probability of reaching the target from the source at a distance $L$ in $n$ steps. The index $G$ denotes the weight of G-searcher path.

The value $z$ is introduced to compare the search models.:
\begin{equation}
    z = \frac{ \mathbb{W}^{G}_{s,t} } {\mathbb{W}^{D}_{s,t}} = \frac{(\mathbb{E}_\varepsilon + 1)\sum_n n  P_{st}(n|L)}{L(1 + \mathbb{E}_{\xi})}
    \label{main_z}
\end{equation}
This is the ratio of the average path weights for each model. The G-searcher wins if the weight of the route is less than the weight of D-searcher, which corresponds to $z<1$. If the weight of the other searcher's route is less, $z>1$. The value in question forms the foundation of the work being undertaken, and an examination will be conducted into how it behaves under different parameters.

On the one hand, this ratio (\ref{main_z}) is entirely controlled by the given uncertainty. For example, if the uncertainty function gives a low probability of a large contribution to the edge's weight, the D-searcher should always win by taking the fewest steps.

On the other hand, if the search is performed on graphs with high connectivity, the G-searcher can win through its strategy of choosing the edge with the minimum weight. However, how many steps does the G-searcher need to take to reach the target? The trade-off occurs between models on the following parameters: the uncertainty function, characterized by the density function $\mathbb{P}_\xi$, and the graph structure, determined by the diameter $D$ and connectivity $c$. In summary, the main question being posed is whether there exists a probability distribution of uncertainty that results in the G-searcher winning.
\begin{equation}
\exists \mathbb{P}_\xi : z < 1
\label{main_question}
\end{equation}
We will show below that such uncertainty exists. Moreover, there are uncertainties under which one of the models will be dominant.

\subsection{\label{sec:mean_first_passage_time_gsearcher}Mean first passage time for G-searcher}
Let us examine formula (\ref{w_greedy}) in detail and introduce the notation
\begin{equation}
m_{st}(L) = \sum_n n  P_{st}(n|L)
\label{our_mfpt}
\end{equation}
As demonstrated in previous studies [\onlinecite{murthy1989mean}, \onlinecite{tishby2022analytical}], this relation represents the mean first passage time (MFPT) for random walks between the source and target. Consider the recurrence equation for the MFPT between arbitrary vertices $i$ and $j$ [\onlinecite{aldous-fill-2014},\onlinecite{papoulis1990random}]:
\begin{equation}
    m_{ij} = 1 + \sum_{l \neq j}^N p_{il} m_{lj}
\label{recurren_mfpt}
\end{equation}
As demonstrated in previous studies [\onlinecite{murthy1989mean}, \onlinecite{tishby2022analytical}], this relation represents the mean first passage time (MFPT) for random walks between the source and target. Consider the recurrence equation for the MFPT between arbitrary vertices $i$ and $j$ [\onlinecite{aldous-fill-2014},\onlinecite{papoulis1990random}]:

\begin{eqnarray}
    m_{st} = 1 & + & \sum_{l \neq j}^{N_1} p_{il} m_{lj}(L - 1) \nonumber \\
               & + & \sum_{l \neq j}^{N_3} p_{il} m_{lj}(L + 1) +  \sum_{l \neq j}^{N_2} p_{il} m_{lj}(L)
    \label{mfpt_3_sums}
\end{eqnarray}

\begin{figure}[b]
\includegraphics[width=0.45\textwidth]{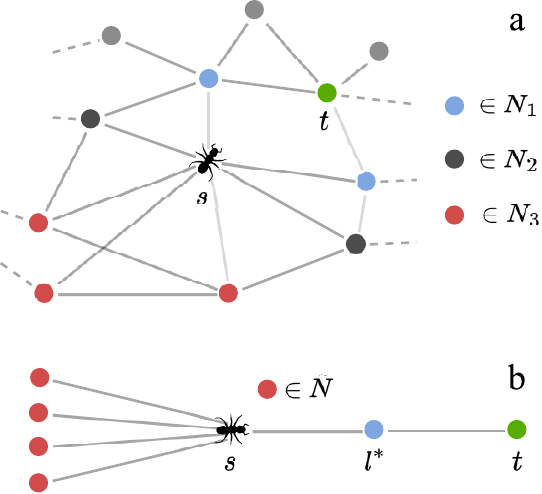}
\caption{\label{fig:e_to_a} (a) An example of splitting adjacent vertices into 3 sets that reduce the distance to the target ($N_1$), do not change it ($N_2$), and increase it ($N_3$). (b) The approximation result for calculating the upper estimate. In which set $N_2$ is included in set $N_1$}
\end{figure}

where the main sum over all adjacent vertices is partitioned into three sets: $N_1$ consists of vertices where the distance to the target decreases upon transition, $N_2$ consists of vertices where the distance does not change, and $N_3$ consists of vertices where the distance increases (see figure \ref{fig:e_to_a}a). Assuming without loss of generality, there exists at least one vertex in the graph where the transition decreases the distance to the target. Combining sets $N_2$ and $N_3$ into $\tilde{N}=N_2\cup N_3$ (refer to figure \ref{fig:e_to_a}b) yields an upper bound for the exact formula (\ref{mfpt_3_sums}):

\begin{eqnarray}
    m_{st} \leq \tilde m_{st}(L) = 1 + \sum_{l \neq j}^{\tilde N} \frac{m_{lj}(L+1) }{c_l} 
    + \frac{m_{l^{*}j}(L-1)}{c_{l^{*}}} 
   \label{mfpt_2_sums}
\end{eqnarray}
where $l^{*}$ is the vertex in $N_1$ that leads to the smallest distance to $t$. 
Assuming the existence of a probability distribution $\mathbb{P}\xi$, which satisfies the condition $z < 1$ for the upper bound $\tilde m_{st}$, the validity of the original assumption (\ref{main_question}) can be established. An accurate expression for the MFPT ($m_{st}$) can then be utilized.
Moreover, upon careful examination of figure \ref{fig:e_to_a}b, the structure of the graph that satisfies this equation can be observed, it is a tree.

In this work [\onlinecite{forster2022exact}], the authors propose a method for calculating the MFPT for random walks on graphs with necklace structure using dimensionality reduction techniques. The authors of the paper claim that every tree can be reduced to a graph with a necklace structure. And they put the case calculation of the MFPT for a $c$-ary tree, which is a tree, each vertex of which (except for leaf vertices) has $c$ child nodes. An exact MFPT equation for a $c$-ary tree between any two vertices is presented by them, which can be utilized here (\ref{main_question}). By utilizing the outcomes from their work, a formula for computing $\tilde m_{st}(L)$ can be obtained:

\begin{equation}
\tilde m_{st}(L) = L \left( \frac{2c^{H+1}}{c-1} - 1\right) - \frac{2 c^{H+1} (1 - c^{-L}) }{(c-1)^2}
\label{mfpt_tree_exact}
\end{equation}
Here, $H$ is the depth of the tree. Moreover, the equation (\ref{mfpt_tree_exact}) can be simplified in the case of highly connected graphs ($c \gg 1$):
\begin{equation}
\tilde m_{st}(L) \approx 2Lc^H
\label{mfpt_tree_approx}
\end{equation}
It is worth noting that in the case of a $c$-ary tree, $c$ is the degree of the terminal vertex, and the difference between these values is one. For highly connected graphs ($c \gg 1$), the values are very close, implying that they are interchangeable. The expression's proof is technically complex; thus, it is included in Supplementary \ref{MFPT_CGC} for simplicity.

\subsection{\label{sec:gsearcher_wins}G-searcher wins prior optimal path}

\begin{figure*}
\includegraphics[width=0.95\textwidth]{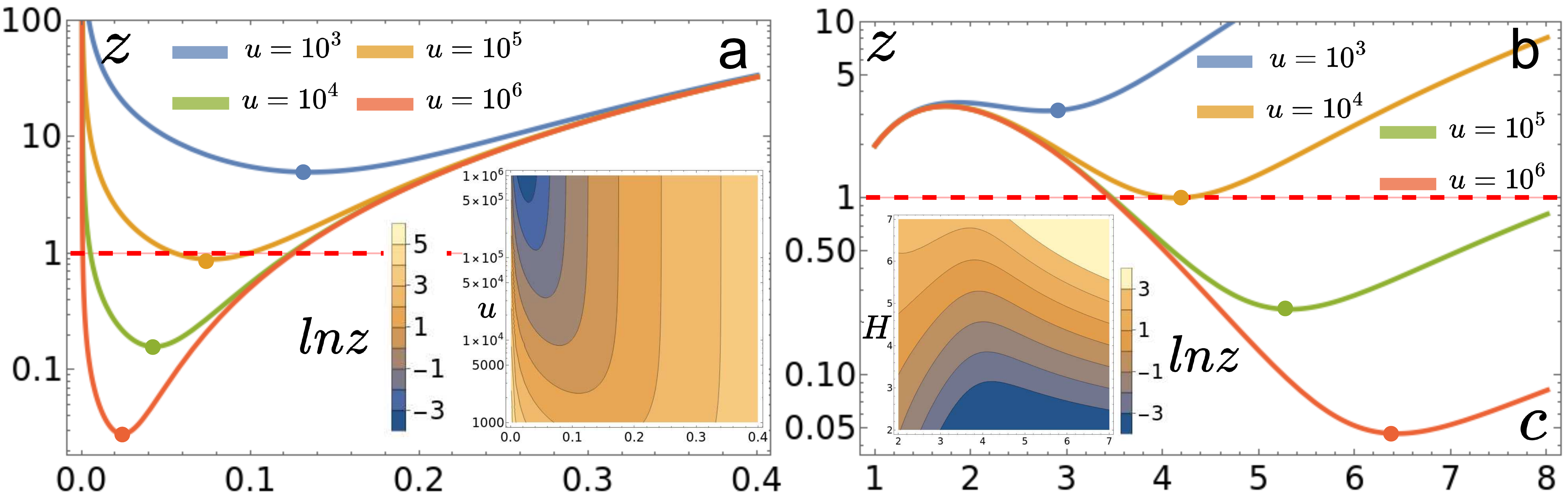}
\caption{\label{fig:res_z_up_z_ch} (a) The main figure shows dependence ${z}$ on $p$, $c = 4, H = 4$. There is the contour plot with critical curves of $\ln z$ depending on $p$ and $u$ on the additional chart in the lower right corner. (b) The main figure shows dependence ${z}$ on $c$ and $\ln z$ depending on $c$ and $H$ on inset figure. The simulation set of uncertainty factors $u = 10^3, 10^4, 10^5, 10^6$ for both main graphs. Parameters for main figure: $p = 0.1, H = 4$, and  for inset figure: $u=10^5$. On both subfigures, the red line represents the division of space into domains $G_w$ (below the line) and $D_w$ (above the line). The colored dots indicate the inflection points corresponding to the probability (\ref{optimal_p}) at which the minimum value of $z$ at fixed parameters $(u,c,H)$ is reached.}
\end{figure*}

After deriving the expression for the MFPT (\ref{mfpt_tree_approx}), the equation of interest (\ref{main_z}) can be solved by substituting the MFPT expression.

\begin{equation}
    \tilde{z} = \frac{(\mathbb{E}_\varepsilon + 1) \tilde m_{v_0 v_L}}{L(1 + \mathbb{E}_{\xi})} = \frac{2c^H(\mathbb{E}_\varepsilon + 1)}{(1 + \mathbb{E}_{\xi})}
    \label{approx_z}
\end{equation}
The resulting equation (\ref{approx_z}) gives an upper estimate of $z$ denoted by $\tilde{z}$, where $z < \tilde{z}$, and $z$ is used later for simplicity. The equation (\ref{approx_z}) reveals that the ratio of expectation weights of paths $z$ is independent of the distance between the source and target $L$ in the case of a large graph.

To answer question (\ref{main_question}) without any loss of generality, the uncertainty model is defined by denoting the extra weight distribution function as a Binomial distribution. The corresponding function $\mathbb{P}_\xi$ is defined as follows:

\begin{align}
\mathbb{P}_\xi(u,p) = 
 \begin{cases}
   p & \text{if } \xi = u \\
   (1-p) & \text{if } \xi = 0
 \end{cases}
\end{align}
Then, $\mathbb{E}_\xi = up$, and $\mathbb{E}_\varepsilon = up^c$ can be easily obtained. Since $\mathbb{P}_\xi(u,p)$ corresponds to a Binomial distribution, the random value $\xi_e$ of the extra weight must take on the value $u$ for all $c$ adjacent edges; otherwise, the G-searcher chooses an edge with less weight. By utilizing these definitions, the equation (\ref{approx_z}) can be converted into (\ref{approx_z_unc}):

\begin{equation}
    z(u,p,c,H) = \frac{2c^H(up^c + 1)}{(up + 1)}
    \label{approx_z_unc}
\end{equation}

The main outcome of the research is the expression presented in (\ref{approx_z_unc}). It is possible to create figures that depend on the uncertainty parameters $z(u,p)$ using this elegant and simple result alone, while keeping the graph unchanged (with constant values of $c$ and $H$),  and vice versa. As shown in figure \ref{fig:res_z_up_z_ch}, these figures indicate that the G-searcher can win. To prove the existence of these regions, the function will be analyzed. Furthermore, the parameter range boundaries where one of the models will always be dominant will be defined below.

\subsection{\label{sec:critical_behavior}Critical behavior}

The figure \ref{fig:res_z_up_z_ch} shows the dependence of ${z}(p)$ at different parameters $u$, while keeping $(c,H)$ fixed. The figure reveals that each parameter $u$ has its optimal value of $p$. To find this optimal probability $p$, the derivative $\frac{\partial z}{\partial p}$ will be analyzed.

The derivative of $z$ with respect to $p$ can be expressed as follows:
\begin{equation}
\frac{\partial{z}}{\partial p} = \frac{2u c^H (up^c(c-1) + c p^{c-1} - 1) }{(up+1)^2}
\label{dzdp}
\end{equation}
The optimal value of $p$ can be obtained by setting the derivative in (\ref{dzdp}) to zero:
\begin{equation}
up^c(c-1) + c p^{c-1} - 1 = 0
\end{equation}
In the limit as $u \gg 1$, the first term in the equation above dominates over the second. The optimal probability $p$ can be expressed as:
\begin{equation}
p(u,c,H) = \big[ u(c-1) \big]^{-\frac{1}{c}}
\label{optimal_p}
\end{equation}
Substituting (\ref{optimal_p}) into the main relation (\ref{approx_z_unc}), the resulting expression is denoted as $z^{*}$:

\begin{equation}
z^*= \cfrac{2c^{H+1}}{(c-1)\left[ u^{1-\frac{1}{c}} \big(c-1\big)^{-\frac{1}{c}} + 1 \right]}
\label{z_uch}
\end{equation}
Next, the equation (\ref{z_uch}) will be investigated in the limit to infinity, and the derivative with respect to $u$ will be found:

\begin{equation}
\lim_{u\to\infty} z^*
 =  \lim_{u\to\infty} \cfrac{2c^{H+1}}{(c-1) \left[ u^{1-\frac{1}{c}}(c-1)^{\frac{1}{c}} + 1 \right]} = 0
 \label{lim_z_u}
\end{equation}

The result of expression (\ref{lim_z_u}) demonstrates that there is an uncertainty value $u$ at which the value of $z$ will be less than 1, indicating that the G-searcher will win. The monotonicity of the function will be investigated to determine any inflection points.

The derivative of $z^*$ with respect to $u$ can be expressed as:

\begin{equation}
\frac{\partial z^*} {\partial u} = -\frac{2 \left(c-1\right)^{-\frac{1}{c}} c^H \sqrt[c]{u}}{\left( u \left(c-1\right)^{-\frac{1}{c}} + \sqrt[c]{u} \right)^2} < 0
\end{equation}

In the found derivative, the numerator and denominator are positive for all values of $u > 0$ and $c \geq 2$. This implies that $z^*$ decreases across the entire interval, eventually tending to zero as $u$ approaches infinity. Thus, for any $c$ and $H$, there exists $u$ and $p$ such that $z<1$, indicating that the G-searcher will outperform the D-searcher. This confirms the main hypothesis (\ref{main_question}).

Next, the function ${z}$ will be investigated with respect to the probability of uncertainty $p$ (see figure \ref{fig:res_z_up_z_ch}a) and the average degree of vertices $c$ (see figure \ref{fig:res_z_up_z_ch}b). The figure \ref{fig:res_z_up_z_ch}b is particularly interesting. As the uncertainty parameter $u$ increases, the function gains additional points of extremum, resulting in convex areas where $z<1$. Consequently, there exists an  $G_w$ where the G-searcher dominates the D-searcher. Conversely, there exists an  $D_w$ where the D-searcher wins under any uncertainty parameters. An attempt can be made to identify a function in the parameter space $(u,p)$ that divides the space into zones with distinct dominant models $D_w$ and $G_w$.

Despite the fact that the analytical search for the boundary between these domains is not analyzed in this paper, limiting relationships can be established that demonstrate the dominance of the D-searcher model under a particular model of uncertainty, regardless of the uncertainty probability $p$ that defines the uncertainty and connectivity $c$ of the graph. It is necessary to investigate the derivative $\frac{\partial z}{\partial c}$.

\begin{equation}
\frac{\partial {z}}{\partial c} = \frac{1}{up+1} (2H c^{H-1} (u p^c + 1) + 2 c^H u p^c \ln p) = 0
\end{equation}

The equality of the derivative $\frac{\partial {z}}{\partial c}$ to 0 yields the following expression:
\begin{equation}
 c = - \frac{H}{\ln p} \left( 1 + \frac{1}{up^c} \right) = - \frac{H}{\ln p} \left( 1 + \frac{e^{(- c \ln p)}}{u} \right)
\label{best_c}
\end{equation}
On the left and right sides of this equation are functions of $c$: $f_1(c) = f_2(c)$. If $f_1$ does not cross $f_2$, then there are no points of extremum on ${z}$, and consequently, there is a region $D_w$. When these functions intersect, there are two points of extremum on ${z}$, but the presence of the region $G_w$ is not guaranteed. The transition point can be found by solving the system of equations:
\begin{align}
\begin{cases}
\frac{\partial f_1(c)}{\partial c} = \frac{\partial f_2(c)}{\partial c} \\
f_1(c) = f_2(c)
\end{cases}
\label{deriv_system}
\end{align}
The first equation of the system yields the expression:
\begin{equation}
1 = \left( - \frac{H e^{(-c \ln p)}}{u \ln p} \right) \left( - \ln p \right)
\end{equation}
By solving for $c$, the result is:
\begin{equation}
c = - \ln \left( \frac{u}{H} \right)/\ln p
\label{c_from_first}
\end{equation}

By inserting the expression (\ref{c_from_first}) into the second equation of the system (\ref{deriv_system}) and applying algebraic techniques, an expression for the critical value of $u_c$ can be derived:

\begin{equation}
u_c = H e^{H+1}
\label{theorem}
\end{equation}

The expression (\ref{theorem}) implies that if $u$ is less than $u_c$, then the function ${z}$ has no extremum points. Consequently, the D-searcher will always win with any combination of parameters $p$ and $c$, regardless of the uncertainty probability $p$ or the connectivity of the graph $c$. This result is noteworthy since it is independent of these factors.

Furthermore, this ratio enables observation of how the a priori weights ($w^p_e = 1$) and additional weights from uncertainty are related. For instance, when the graph height $H=4$, the critical additional weight of uncertainty is approximately $ u_c \approx 6*10^2$. However, when the graph height $H=9$, the critical additional weight of uncertainty is approximately $ u_c \approx 2*10^5$. Below, we will confirm this statement through numerical calculations on graphs with varying structures.

\section{\label{sec:results}Results}
This section presents simulation results of both searcher models on graphs with different structures. Subsection \ref{sec:synt_graphs} presents simulation results on synthetic graphs, such as random regular, n-dimensional uniform lattices, Small-world, and Erd\H{o}s-R\'enyi graphs, in comparison with theoretical estimates. Subsection \ref{sec:real_graphs} explores several real graphs, including a graph of friends in a social network based on anonymized Facebook data and a graph of the vessel system of the mouse brain.  Scale-free and Barab\'asi–Albert graphs are singled out for separate study in subsection \ref{sec:robustness}, which are characterized by a heavy-tailed connectivity distribution, and their robustness to global search models is analyzed. In studying all graphs, the simulation results are compared with theoretical ones (\ref{approx_z_unc}): dependencies $z(c,H)$ for fixed values of $u,p$ and $z(u,p)$ for fixed values of $(c, H)$. The height of the graph, related to the diameter by the relation $D = 2H$, is used for the graphs discussed in this section.

\subsection{\label{sec:synt_graphs}Comparison with numerical simulation on different random graph models}

The process of obtaining the experimental dependencies $z(u,p)$ with fixed values of $(c,H)$ and $z(c,H)$ with fixed values of $u,p$ is the same and based on a similar algorithm.

The source and target are randomly selected from the graph at the beginning of the simulation. The shortest distance between them is found using any global search algorithm. 
Then, the edge weights are sampled from the $\mathbb{P}\xi(u,p)$ distribution and summed up to obtain the weight of the route for the D-searcher. Next, a G-searcher is launched from the source, which only walks along edges with the minimum weight. It's important to note that during the D-searcher walk, the weights for the edges are sampled from the same uncertainty distribution $\mathbb{P}\xi(u,p)$ at each step, even if the vertex or edge has already been visited. Once the G-searcher reaches the target vertex, the weight of the route for the G-searcher is obtained. This process is repeated 1000 times for each set of parameters, and the results are averaged.

The simulation algorithm for the $z(u,p)$ and $z(c,H)$ dependencies is presented in Supplementary \ref{sim_alg}.\ref{alg:z_up} and Supplementary \ref{sim_alg}.\ref{alg:z_cd}, respectively. The basic simulation algorithm is described in Supplementary \ref{sim_alg}.\ref{alg:simulation}.

As random graphs are utilized in this section, a brute-force search of the generation parameters was performed to obtain a graph with specific parameters. The generation parameters for each graph model are described separately. The Python package \textit{networkx} is utilized for all operations on graphs [\onlinecite{hagberg2008exploring}]. However, for certain types of graphs, creating a graph with specific values of $(c,H)$ necessitates substantial computing resources. Hence, for these types of graphs, only the $z(u,p)$ dependency with fixed values of $(c,H)$ is modeled and compared.

\subsubsection{\label{sec:random_regular_graphs}Random regular graphs}

\begin{figure}[t]
\includegraphics[width=0.45\textwidth]{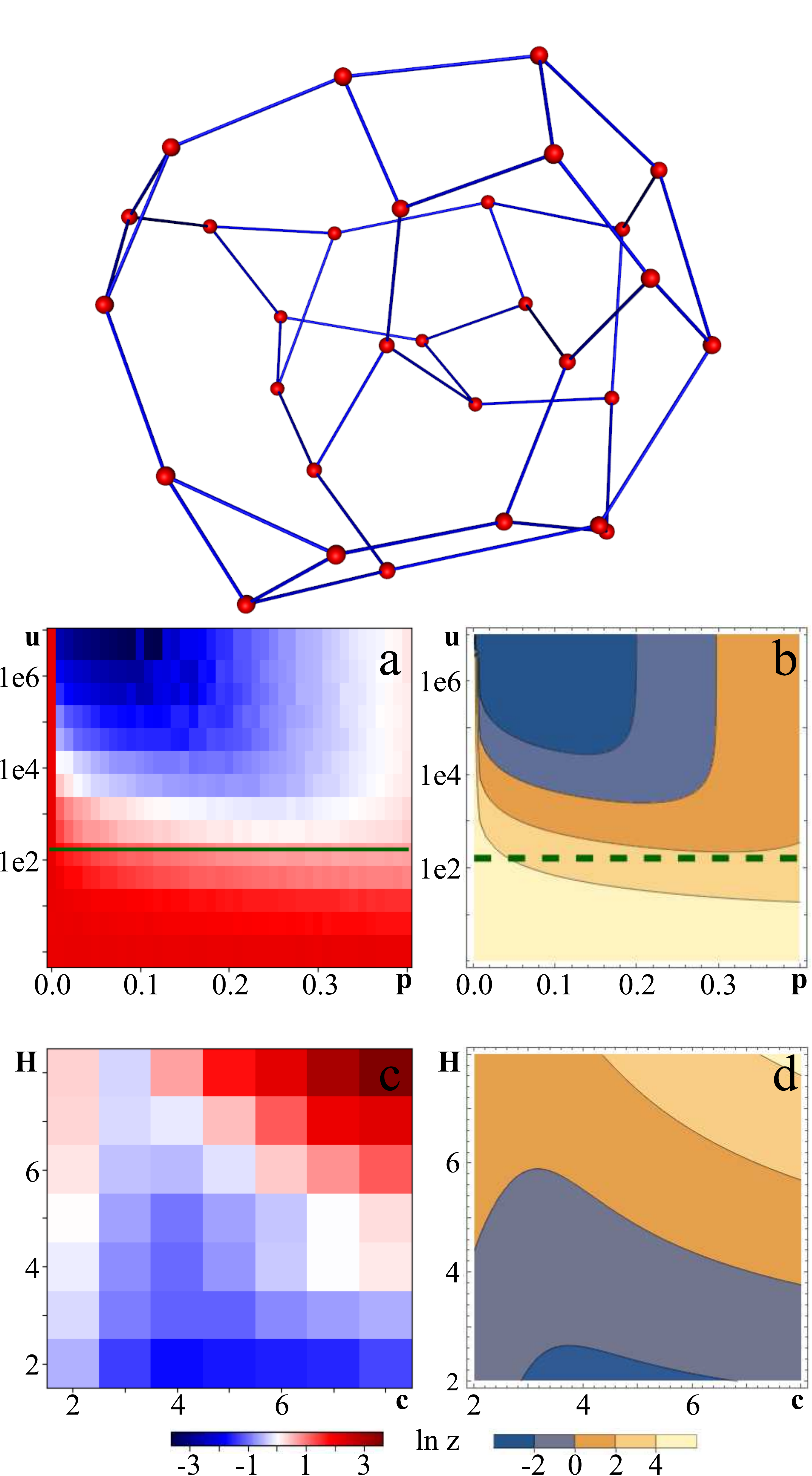}
\caption{\label{fig:regular_results} The figure shows the results of the simulation (a,c) on random regular graphs with theoretical (b,d) ratios (\ref{approx_z_unc}): $\ln z$ on $(c,H)$ with fixed $u=500$ and $p=0.15$ on the figures (c, d), $\ln z$ on $u,p$ with fixed regular graph $c=6$ and $H=3$ showed on subfigures a, b. The figures a, b show a green line corresponds to the critical value $u_{crit}$ (\ref{theorem}). The area below this line is belonging to region $D_w$.}
\end{figure}

The analysis of algorithms commonly employs random $c$-regular graphs as the primary type of graphs [\onlinecite{janson2011random}, \onlinecite{wormald1999models}]. A regular graph is defined as a graph in which every vertex has the same number of neighbors $c$. One needs to specify the desired degree of vertices and the number of total vertices to generate a regular graph using the python package \textit{networkx}. The number of vertices is determined to obtain the required graph diameter.

Figure \ref{fig:regular_results} shows a regular graph in the upper area as a demonstration. Subfigures \ref{fig:regular_results}a and \ref{fig:regular_results}c illustrate the simulation results for dependencies $\ln z(u,p)$, where $c = 6$, and $H=3$ and $\ln z(c,H)$ for a fixed uncertainty model where $u = 500$, and $p=0.15$, respectively. Each grid cell of the simulation results was obtained by averaging over the ensemble of runs with fixed parameters $(u,p,c,H)$, which corresponds to $\mathbb{E}[z(u,p,c,H)]$. The region dominated by the D-searcher is represented by the red area, whereas the blue area is dominated by the G-searcher, and the transition region is white on experimental subfigures (\ref{fig:regular_results}a, \ref{fig:regular_results}c) or the critical curve on theoretical results (\ref{fig:regular_results}b, \ref{fig:regular_results}d).

Subfigures \ref{fig:regular_results}b and \ref{fig:regular_results}d show the corresponding theoretical dependencies in the form of contour plots calculated with the same parameters as the simulation. On the theoretical results graph, a separate contour highlights the transition area between the dominant search engines. This contour between light blue (G-searcher dominates) and orange (D-searcher dominates) corresponds to the expression $\ln z = 0$. As shown in subfigure \ref{fig:regular_results}d, this contour has a bell shape with these parameters.

Comparing the simulation results from left to right with the theoretical predictions, it is observed that the former qualitatively and quantitatively match the latter. The transition contour between optimal searcher models, predicted from theoretical calculations, is repeated in both simulation and theoretical results. In addition, figures \ref{fig:regular_results}a and \ref{fig:regular_results}b mark the value $u_c$ with a green line, which cuts off the region where the D-searcher always predominates. This queue is clearly visible in the figure \ref{fig:regular_results}a with numerical simulation. The green line runs along the bottom edge of the transition region, and there are no blue areas corresponding to the dominance of the G-searcher in the region below the green line.

Since the degree of all vertices is constant, the simulation results on this type of graph closely match the theoretical ones. A different colormap is used to reflect numerical results and theoretical estimates.

\subsubsection{\label{sec:n_dimensional_uniform_lattice}N-dimensional uniform lattice}

A lattice graph is a graph whose drawing, embedded in some Euclidean space, forms a regular tiling. This type of graph may more shortly be called just a lattice, mesh, or grid. The size, grid dimension, and periodicity flag are used for generation. Unfortunately, generating graphs with all the necessary parameters $(c,H)$ for this type of graph is difficult. Therefore, the analysis is confined to considering the behavior of the quantity $z$ on parameters $(u,p)$ for a fixed uniform lattice graph.

\begin{figure}[t]
\includegraphics[width=0.45\textwidth]{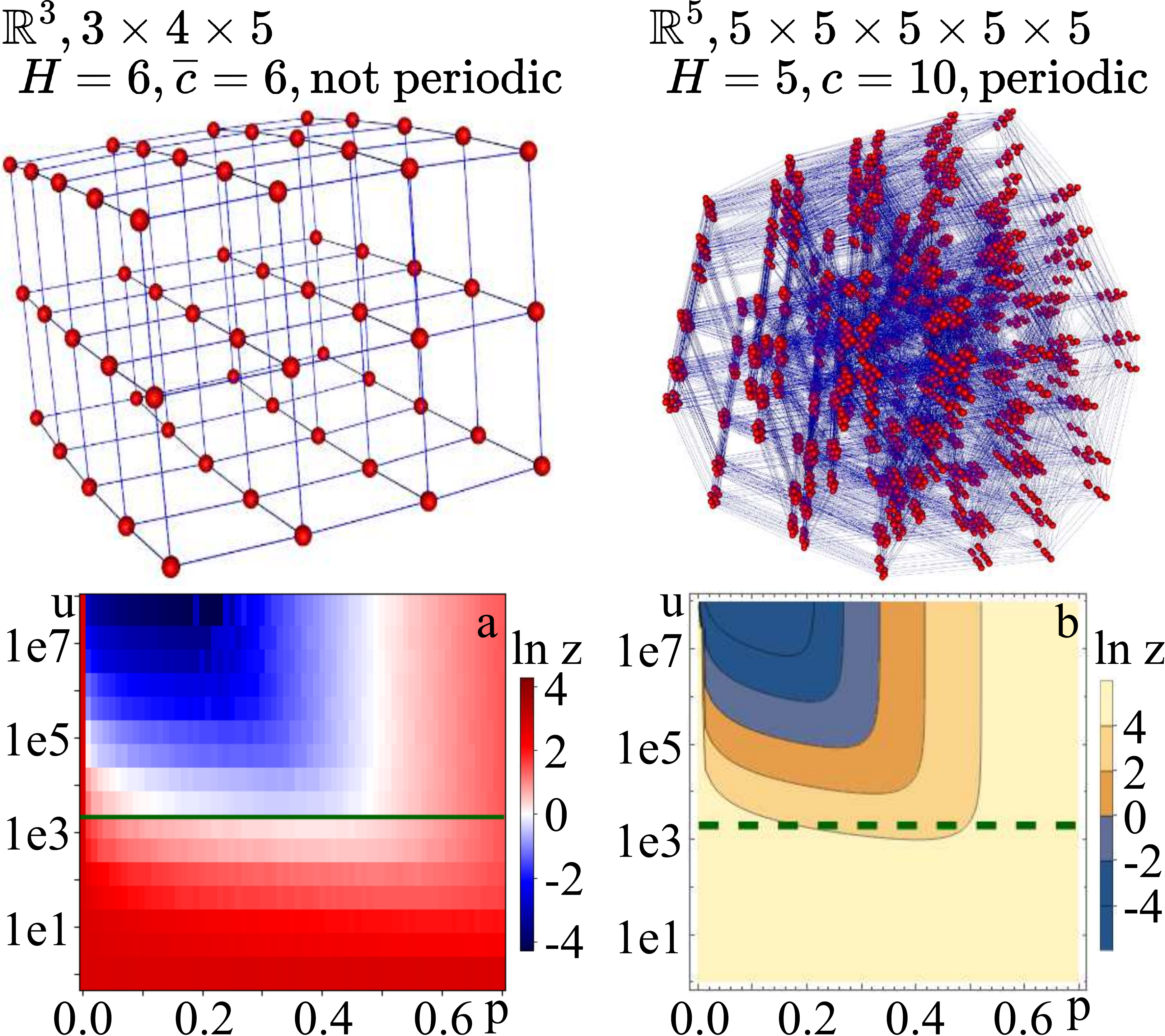}
\caption{\label{fig:grid_results} The figure shows the results of the simulation (a) on graphs with theoretical (b) ratios (\ref{approx_z_unc}): $\ln z$ on $u,p$ with fixed 5-dimension lattice graph ($c=10$ and $H=5$ showed in the upper right corner) at the bottom. The figures on the bottom show a green line corresponds to the critical value $u_{crit}$ (\ref{theorem}). The area below this line is belonging to region $D_w$.}
\end{figure}

In Figure \ref{fig:grid_results}, the upper left side shows a three-dimensional uniform lattice ($\mathbb{R}^3,3\times4\times5$) with 3 vertexes along the X axis, 4 vertexes along the Y axis, and 5 vertexes along the Z axis. This lattice is not periodic, that is, the extremity vertices are not connected to each other. As a consequence, the degree for the outer vertices is less than the degree of the inner vertices.

In the same figure, the upper right part demonstrates a 5-dimensional periodic lattice used for numerical simulation. Unlike a non-periodic lattice, all vertices of such a graph have the same degree value. Moreover, when the lattice dimension increases by one, two edges are added for each vertex. These considerations lead to the following dependence $\mathbb{R}^n, c=2^n$ for an n-dimensional uniform lattice. As a consequence, specifically for this graph, the degree of each of its vertices has the value $c=10$.

Figures \ref{fig:grid_results}a and \ref{fig:grid_results}b show the numerical and theoretical results of dependence $\ln z(u,p)$ on a fixed graph, respectively. When comparing subfigures \ref{fig:grid_results}a and \ref{fig:grid_results}b, there is a numerical and qualitative agreement between the theoretical results and the simulation results. In addition, the green line, which indicates the critical value of the uncertainty function $u_c$ below which global search algorithms dominate, also agrees well with the results obtained and passes just below the inflection line. The area below this line belongs to region $D_w$.

\subsubsection{\label{sec:small_world_networks}Small-world networks}

A small-world network is a type of graph where most vertices are not neighbors but are at a small distance from each other [\onlinecite{watts1998collective}], i.e. $L \propto \ln N$, where L is the distance between two random vertices from the graph, and $N$ is the count of vertices in the graph.

\begin{figure}[t]
\includegraphics[width=0.45\textwidth]{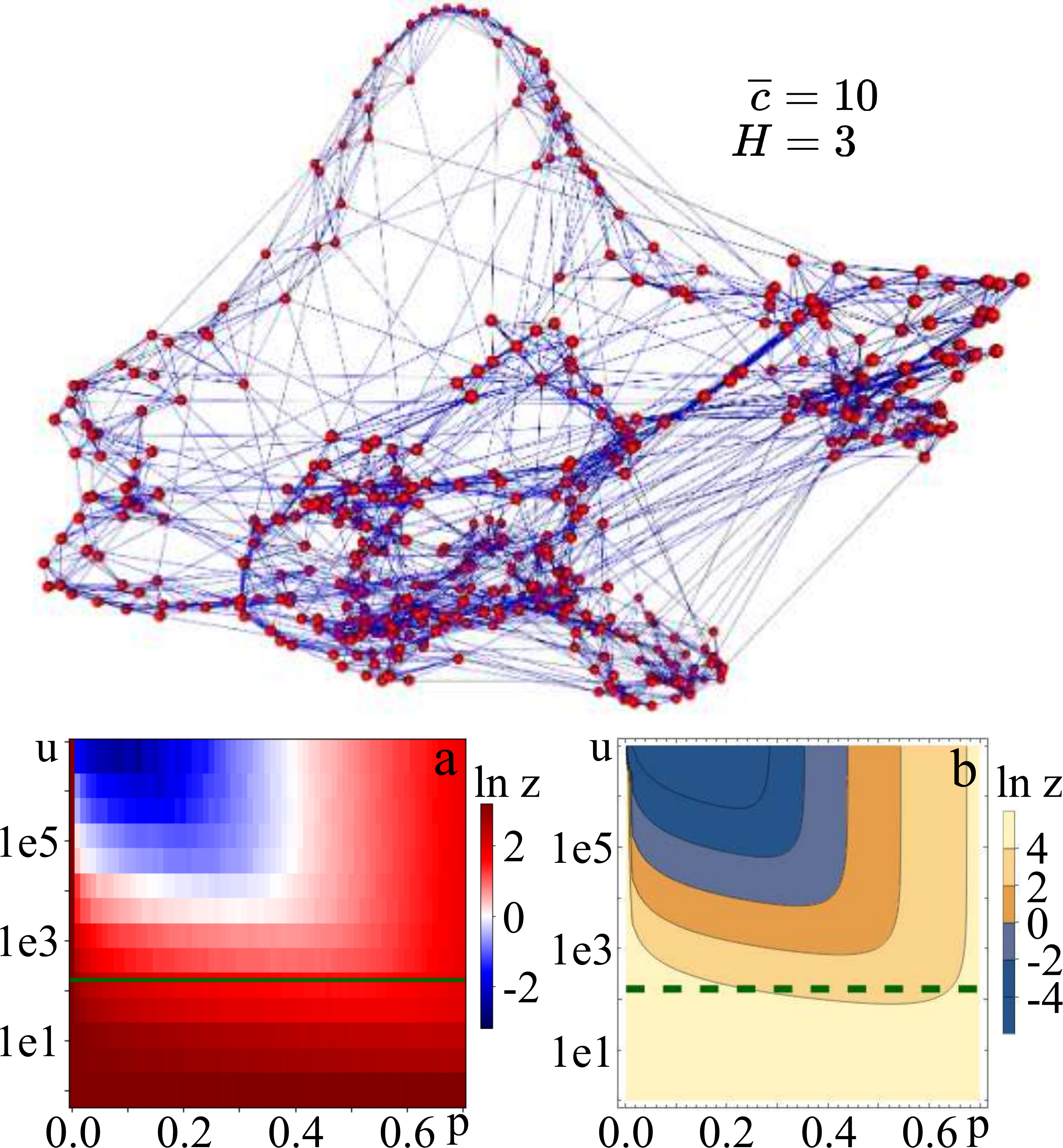}
\caption{\label{fig:small_world_results1}  The figure shows the results of the simulation (a) on Small-world graphs with theoretical(b) ratios (\ref{approx_z_unc}) figures in the bottom represent dependence of $\ln z$ on $u,p$ with fixed Small-world structure graph ($\overline{c}=10$ and $H=3$), which is shown on the top. The graphs on the right show a green line representing critical value $u_{crit}$ (\ref{theorem}). The area below this line is belonging to region $D_w$.}
\end{figure}

The Watts-Strogatz model [\onlinecite{watts1998collective}] is used to generate a graph for numerical simulations. This model involves a ring of $N$ vertices (a ring topology), each connected to the nearest $k$ neighbors. The edges' input or output vertex is then changed with some probability $p$, causing the degree of one vertex to decrease while the other increases by 1. Thus, the average degree of the vertices remains constant, but the vertex degree distribution becomes wider as the probability $p$ increases.

This type of graph has great potential, allowing for a range between regular and random graphs. A Small-world graph was built using the Watts-Strogatz model with 5000 vertices, each with 10 neighboring vertices in a ring topology, and a rewiring probability of 0.5. Generating graphs for the entire set of parameters $(c,H)$ is computationally demanding, so the study will be restricted to the dependence $\ln z(u,p)$ for a fixed graph. Figure \ref{fig:small_world_results1} shows the Small-world graph used for the calculation with parameters $\overline{c}=10$ and $H=5$.

Figure \ref{fig:small_world_results1}a shows the simulation results, and Figure \ref{fig:small_world_results1}b shows the theoretically predicted dependence $\ln z(u,p)$. The blue region indicates the area where the G-searcher is more stable, while the red area characterizes the stability of the D-searcher. The theoretically estimated dependency is shown in contour view, with the region dominated by the D-searcher indicated in shades of orange. Similar to the previous types of graphs, visual analysis shows good agreement between theoretical estimates and numerical simulation results. The critical curve clearly cuts off the region where global search is stable.

\subsubsection{\label{sec:erdos_renyi}Erd\H{o}s-R\'enyi graphs}

Another widely used model for generating random graphs is the Erd\H{o}s-R\'enyi model [\onlinecite{erdos1960evolution}]. This model assumes that edges are independent and equally likely, which is not suitable for modeling many real-life phenomena. As a result, the degree distribution of Erd\H{o}s-R\'enyi graphs is characterized by a lack of heavy tails, while real systems often exhibit power-law degree distributions. Additionally, Erd\H{o}s-R\'enyi graphs lack two important properties commonly observed in real networks [\onlinecite{ravasz2002hierarchical}]: they have a low clustering coefficient and do not account for vertex formation, resulting in a degree distribution that converges to a Poisson distribution instead of a power law.

\begin{figure}[b]
\includegraphics[width=0.45\textwidth]{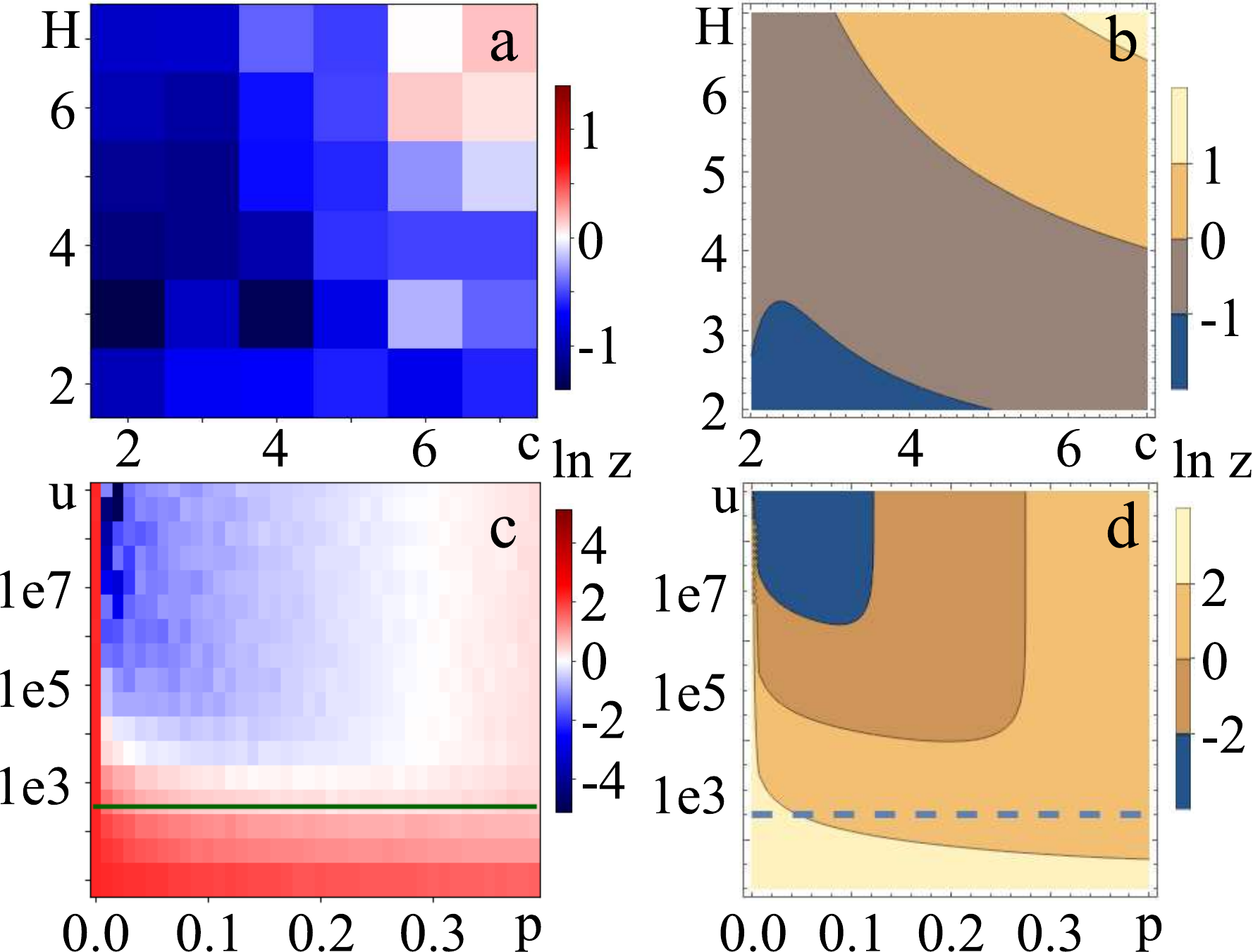}
\caption{\label{fig:erdos_renyi_result}  The figure shows the results of the simulation (a, c) on Erd\H{o}s-R\'enyi graphs with theoretical (b, d) ratios (\ref{approx_z_unc}): $\ln z$ on $(c,H)$ with fixed $u=10^4$ and $p=0.01$ on the top (a, b), $\ln z$ on $u,p$ with fixed Erd\H{o}s-R\'enyi graph ($\overline{c}=7$ and $H=4$) on the bottom (c,d). The graphs on the right show a green line corresponds to the critical value $u_{crit}$ (\ref{theorem}). The area below this line is belonging to region $D_w$.}
\end{figure}

\begin{figure*}[t]
\includegraphics[width=1\textwidth]{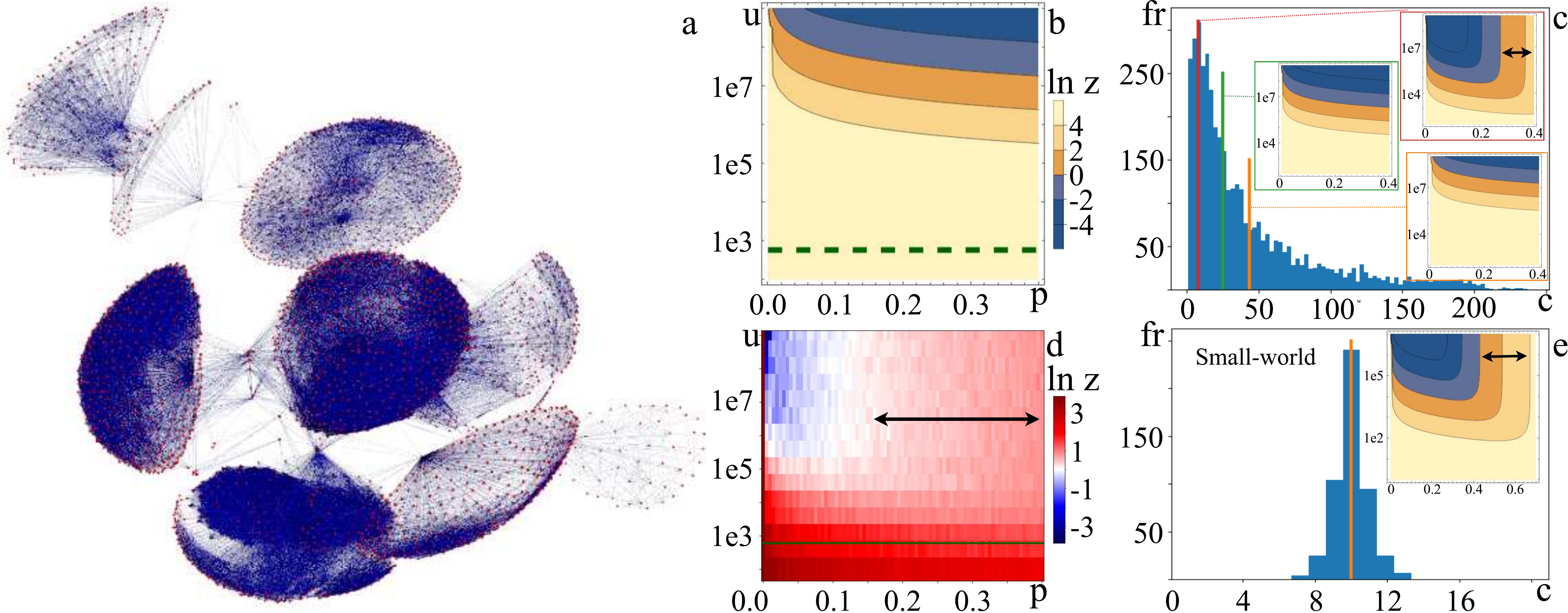}
\caption{\label{fig:facebook} The figure shows the results of the simulation (d) on Facebook graph with theoretical (b) ratios (\ref{approx_z_unc}). Subfigures in the middle (b,d) represent dependence of $\ln z$ on $u,p$ on Facebook graph ($H=3, \overline{c}=43$), which is shown on the left (a). The green line that represents dependency (\ref{theorem}). The subfigure (c) shows the vertex degree distribution for Facebook graph. The mean, median, and maximum of the distribution are marked. For all of them, theoretical estimates are constructed on insets. The subfigure  (e) shows the vertex degree distribution for the Small-world structure graph.}
\end{figure*}

Despite these limitations, as shown in Figure \ref{fig:erdos_renyi_result}, this model agrees well with theory. Similar to the previous models, the numerical simulation results (Figure \ref{fig:erdos_renyi_result}a and \ref{fig:erdos_renyi_result}c) closely match the theoretical predictions (Figure \ref{fig:erdos_renyi_result}b and \ref{fig:erdos_renyi_result}d). Furthermore, the generated graphs exhibit a narrow bell-shaped degree distribution. Notably, this model's parameter space $(c,H)$ can be easily varied, allowing for the generation of multiple graphs for numerical simulations of $\ln z(c,H)$. Figures \ref{fig:erdos_renyi_result}a and \ref{fig:erdos_renyi_result}b show numerical and theoretical results with a fixed uncertainty model $u=10^4, p=0.01$, while Figures \ref{fig:erdos_renyi_result}c and \ref{fig:erdos_renyi_result}d display calculations for a fixed graph with parameters $\overline{c}=7$ and $H=4$.

Moreover, subfigure \ref{fig:erdos_renyi_result}b illustrates that the critical curve has a monotonic shape, in contrast to the bell shape observed when simulating on a regular graph (see Figures \ref{fig:regular_results}c and \ref{fig:regular_results}d).

\subsection{\label{sec:real_graphs}Uncertainty in real world networks}
In the preceding section, only synthetic graphs of specific models were studied, allowing the generation of graphs for the numerical calculation of $\ln z(c,H)$ for just two graph models (Erd\H{o}s-R\'enyi, random regular). This section aims to demonstrate and discuss the results of a numerical experiment on a real graph and focus on the limits of applicability of the theoretical calculations.

\subsubsection{\label{sec:facebook_graph}Facebook graph}
Consider the anonymized Facebook friends list which was obtained from an open source dataset [\onlinecite{facebook_dataset}]. You can see the graph in more detail at the link in the supplemental materials [\onlinecite{facebook.2023}]. Subfigure \ref{fig:facebook}a displays the Facebook graph that was used in the numerical simulation. The community structure is clearly visible in this figure. Figures \ref{fig:facebook}b and \ref{fig:facebook}d exhibit the results of numerical simulation and theoretical calculations on the Facebook graph with fixed parameters $H = 3$ and $\overline{c}=43$. A significant deviation of the experimental critical curve $\ln z = 0$ from the theoretical curve, constructed based on the average vertex degree in the graph, can be observed. This deviation is assumed to be related directly to the distribution of vertex degrees. Thus, the graph structure between communities has the greatest impact on the shape of the resulting stable region. Subfigure \ref{fig:facebook}c displays the change in the shape of the critical curve if the median or maximum degree (the degree of the most frequently occurring peak) is used for the evaluation instead of the mean degree. By reducing the average degree of vertices used in the theory, the shape of the curve becomes more similar to the experimental one and the stability region of the global search increases (black double arrow, see figures \ref{fig:facebook}c, \ref{fig:facebook}d, \ref{fig:facebook}e). However, using the average degree of vertices works well for grid and Small-world graphs since their degree distribution is relatively narrow, and for periodic grids, it is constant. Moreover, in \ref{fig:facebook}d, there are no areas below the green line where G-searcher dominates, which is consistent with theoretical predictions.

\subsubsection{\label{sec:brain_vessel_graphs}Brain Vessel Graphs}

\begin{figure*}[t]
\includegraphics[width=1\textwidth]{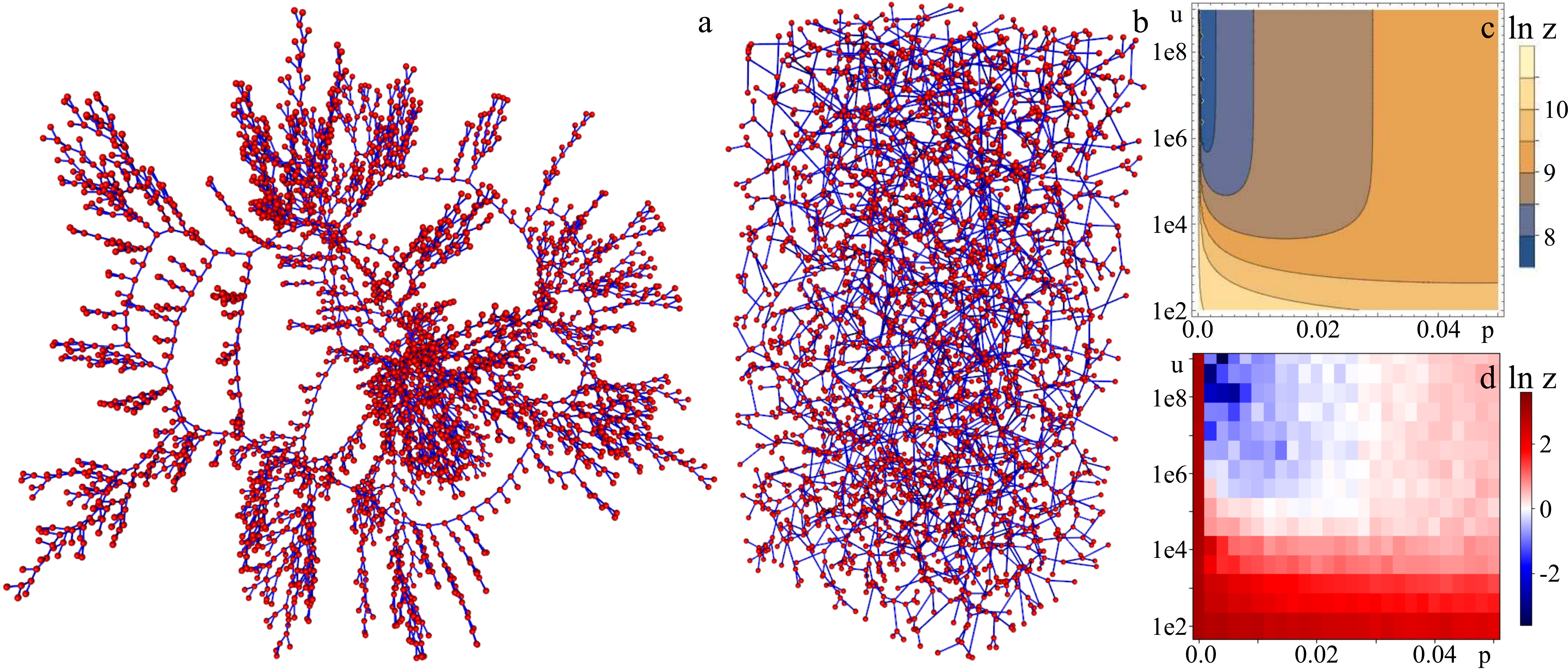}
\caption{\label{fig:vessel}  The figure shows the results of the simulation (c) on brain vessel graph ($\overline{c}=2$ and $H=34$) with theoretical (d) ratios (\ref{approx_z_unc}) $\ln z$ on $u,p$. The subfigure (b) shows the input graph built on the basis of the vascular system of the brain with the position of the vertices of the graph preserved. The subfigure (a) shows the same graph with vertex positions changed to better reflect the topology of the graph.}
\end{figure*}

Figure \ref{fig:vessel} shows simulation results for the extracted spatial vessel graph. This test is based on the work on creating a neural network and collecting a dataset for training it to generate a graph of the vascular system of the brain [\onlinecite{schneider2012tissue}]. You can see the graph in more detail at the link in the supplemental materials [\onlinecite{vessel.2023}]. Using a graph representing the whole entire structure of the circulatory system is not possible, since this graph has a million vertices and several times more connections. Therefore, a subgraph was extracted based on a rectangle given in space (see figure \ref{fig:vessel}b). If one of the vertices of a connection is a vertex not included in the subgraph, such a connection is removed.
As can be seen from the results, the theory in figure \ref{fig:vessel}c has quantitative discrepancies with the simulation results in figure \ref{fig:vessel}d. However, the shape of the critical curve qualitatively coincides with the theory. Moreover, the theoretical cutoff $u_c$ (\ref{theorem}) is located at a level approximately in the region $u \approx 10^{16.7}$, since $H=34$ for this graph. And it is located further behind the transition region under consideration. Which, in turn, does not correspond to theoretical conclusions, since there is a region where G-searcher is stable.

Similar to the previous case with a social network graph case (Facebook graph), this discrepancy is attributed to a specific distribution of vertex degrees.
For a better representation, the location of the graph nodes in space has been transformed. The original graph is shown in figure \ref{fig:vessel}b, and the transformed one is shown in figure \ref{fig:vessel}a.
Almost half of the vertices have degree 1, the other half has degree 3, and only a small part of the vertices have degree 2. Thus, the average degree of the vertices is $\overline{c} \approx 2.002$. This value is assumed to be very small for the applicability of the theory, since the condition about working with high connectivity graphs $c \gg 1$ was added to the final expression (\ref{mfpt_tree_approx}) during the derivation of the model. Thus, the final value in the expression for MFPT begins to be strongly influenced by the second term. This may be the reason for the quantitative discrepancies between the numerical results and the theoretically predicted ones.

\subsection{\label{sec:robustness}Robustness of Scale-free networks}
The previous section considered several real graphs where the numerical simulation results were quantitatively different from the theoretical ones. It is reasonable to assume that this is due to the distribution of vertex degrees in the graph. Therefore, this section will consider a certain type of graph called Scale-free, which has a heavy-tailed distribution of vertex degrees. As a consequence, the limits of applicability of the theory will be studied in more detail.

\subsubsection{\label{sec:scale_free_graphs}Scale-free graphs}
The Directed Scale-Free Graphs model [\onlinecite{bollobas2003directed}] proposes the generation of directed scale-free graphs that grow by attaching edges to vertices with the highest in or out degree. This model gives a power-law distribution of vertex degrees consistent with those seen on the World Wide Web. Creating a graph requires setting the following parameters:
\begin{itemize}
\item Number of vertices
\item The probability of adding a new vertex connected to an existing vertex chosen at random according to the degree distribution.
\item The probability of adding an edge between two existing vertices. One existing vertex is selected randomly according to the distribution of incoming degrees, and the other is selected randomly according to the distribution of outgoing degrees.
\item The probability of adding a new vertex connected to an existing vertex picked at random according to the distribution of out-degrees.
\end{itemize}
It is also important to note that this type of generation allows for the creation of a set of graphs that enable an experimental dependency $z(c,H)$ to be obtained.

\begin{figure}[t]
\includegraphics[width=0.5\textwidth]{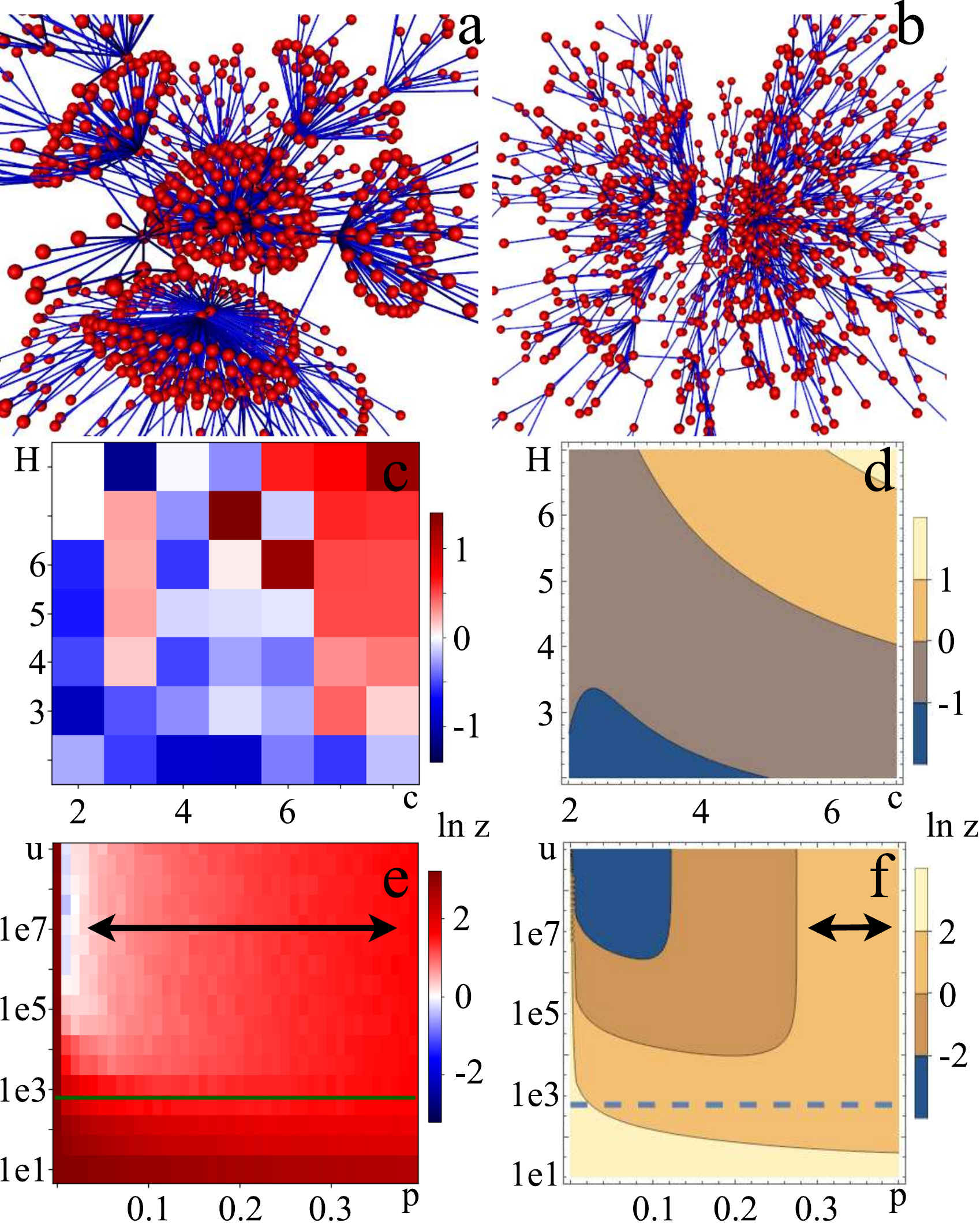}
\caption{\label{fig:directed_scale_free_graph_results} The figure shows the results of the simulation (a,c) on graphs with theoretical (b,d) ratios (\ref{approx_z_unc}): $\ln z$ on $(c,H)$ with fixed $u=10^4$ and $p=0.01$ on subfigures (a, b), $\ln z$ on $u,p$ with fixed the Directed scale-free graph ($\overline{c}=8$ and $H=4$)  on subfigures (c, d). Example graphs at the top of the figure. The subgraphs (c,d)  show a green line corresponds to the critical value $u_{crit}$ (\ref{theorem}). The area below this line is belonging to region $D_w$.}
\end{figure}

Figures \ref{fig:directed_scale_free_graph_results}a and \ref{fig:directed_scale_free_graph_results}b display the structure of Scale-free graphs. For fixed uncertainty parameters $u=10^4, p=0.01$, figures \ref{fig:directed_scale_free_graph_results}c and \ref{fig:directed_scale_free_graph_results}d illustrate the results of numerical simulation and theory for dependence $\ln z(c,H)$. The theory predicts a monotonic critical curve, but the numerical results (subfigure \ref{fig:directed_scale_free_graph_results}c) do not qualitatively distinguish this curve.

For a fixed graph with parameters $\overline{c}=8,H = 4$, figures \ref{fig:directed_scale_free_graph_results}e and \ref{fig:directed_scale_free_graph_results}f demonstrate the results of numerical simulation and theory for dependence $\ln z(u,p)$. The numerical results have a wider region with a stable global search than theoretically predicted, but the green line excludes the parameters under which the G-searcher cannot dominate. This agrees well with experimental results. Similar behavior was observed with the Facebook graph (see fig. \ref{fig:facebook}): a narrowing region with the dominance of the G-searcher, but with a critical value $u_c$ that agrees well with the experiment.

\subsubsection{\label{sec:barabasi_albert}Barab\'asi–Albert model} 

An additional model exists for generating networks without using Barab'asi–Albert's scale models [\onlinecite{barabasi1999emergence}]. The Barab'asi–Albert model proposes an explanation for the phenomenon where new pages on the World Wide Web tend to link to the most popular existing pages (hubs), like Google. When selecting a new page to link to, a specific page is chosen at random, with the likelihood of selection being proportional to the degree of the page. According to the Barab'asi–Albert model, this accounts for the preferred attachment probability rule.

\begin{figure}[t]
\includegraphics[width=0.5\textwidth]{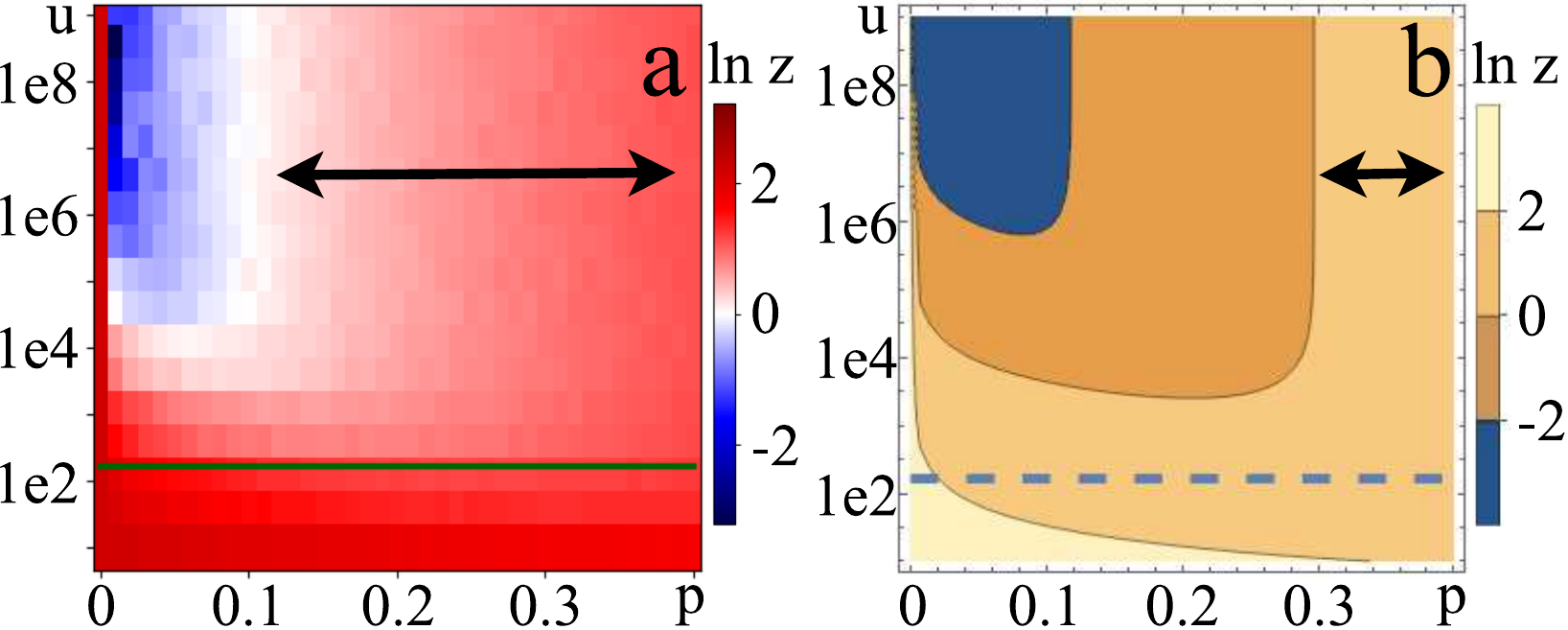}
\caption{\label{fig:barabasi_albert_results} The figure shows the results of the simulation (a) on graphs with theoretical (b) ratios (\ref{approx_z_unc}) $\ln z$ on $u,p$ with fixed Barab\'asi–Albert graph ($\overline{c}=6$ and $H=3$). The both subfigures show a green line corresponds to the critical value $u_{crit}$ (\ref{theorem}). The area below this line is belonging to region $D_w$.}
\end{figure}

A preferential attachment mechanism is utilized in this model. Statistical similarities have been observed between graphs generated by this model and real graphs created by humans, such as the World Wide Web, the citation network, and social networks [\onlinecite{amaral2000classes}]. Such networks are regarded as scale-free and consist of specific vertices known as hubs, which have a high degree.

Regrettably, generating a set of graphs for calculating dependence $z(c,H)$ with this model necessitates significant computational resources. Consequently, the analysis was limited to examining the dependence $\ln z(u,p)$ on the generated graph with parameters $\overline{c}=6,H=3$, as shown in figure \ref{fig:barabasi_albert_results}. As demonstrated in the figure, a similar scenario was observed, where the stable D-searcher region exceeded the theoretical prediction (black double arrow, see figures \ref{fig:barabasi_albert_results}). Meanwhile, the results of numerical simulation displayed qualitative agreement with the theoretically predicted outcomes, and the experimental results corresponded to the predicted critical value of the uncertainty model parameter $u_c$.

This section presents the simulation results compared to the theoretically predicted dependences for G-searcher and D-searcher on different synthetic graph models, along with real social network graphs and the vessel system in the mouse brain. Cases were identified where the simulation results diverged from the theory, primarily due to the distribution of vertex degrees in the graph. The theory operates best on graphs where the width of this distribution is narrow. To test this assumption, several types of synthetic data with a heavy-tailed distribution for degrees of vertices were examined, discovering that the theory only qualitatively described the behavior of the searcher for such cases, exhibiting quantitative discrepancies. This prompts further refinement of the model.

\section{\label{sec:discussion}Discussion}

In this section, we present ideas for future research and potential practical applications of the developed model and its predictions, each of which deserves in-depth investigation.

Our model has practical applications in search algorithms on graphs in unpredictable and dynamic environments, where agents must make decisions with incomplete information. Furthermore, analyzing network growth and evolution to maintain stable search properties can aid in designing and optimizing network architecture, a promising task with numerous applications. In addition to digital algorithms, the behavior of biological agents in natural environments under uncertain conditions is intriguing. In the final part of this section, we provide observations on the relationship between our proposed model and the evolutionary behavior of miniature biological species, particularly insects.

It is worth noting that the practical implications of our model extend beyond the examples mentioned above. The potential applications are vast and varied, and we are excited to explore them further in future research.

\subsection{\label{sec:transition_critical_curve} Transition of the Critical Curve from Monotonic to Bell-shaped and Stability of Network Growth}

\begin{figure}[t]
\includegraphics[width=0.4\textwidth]{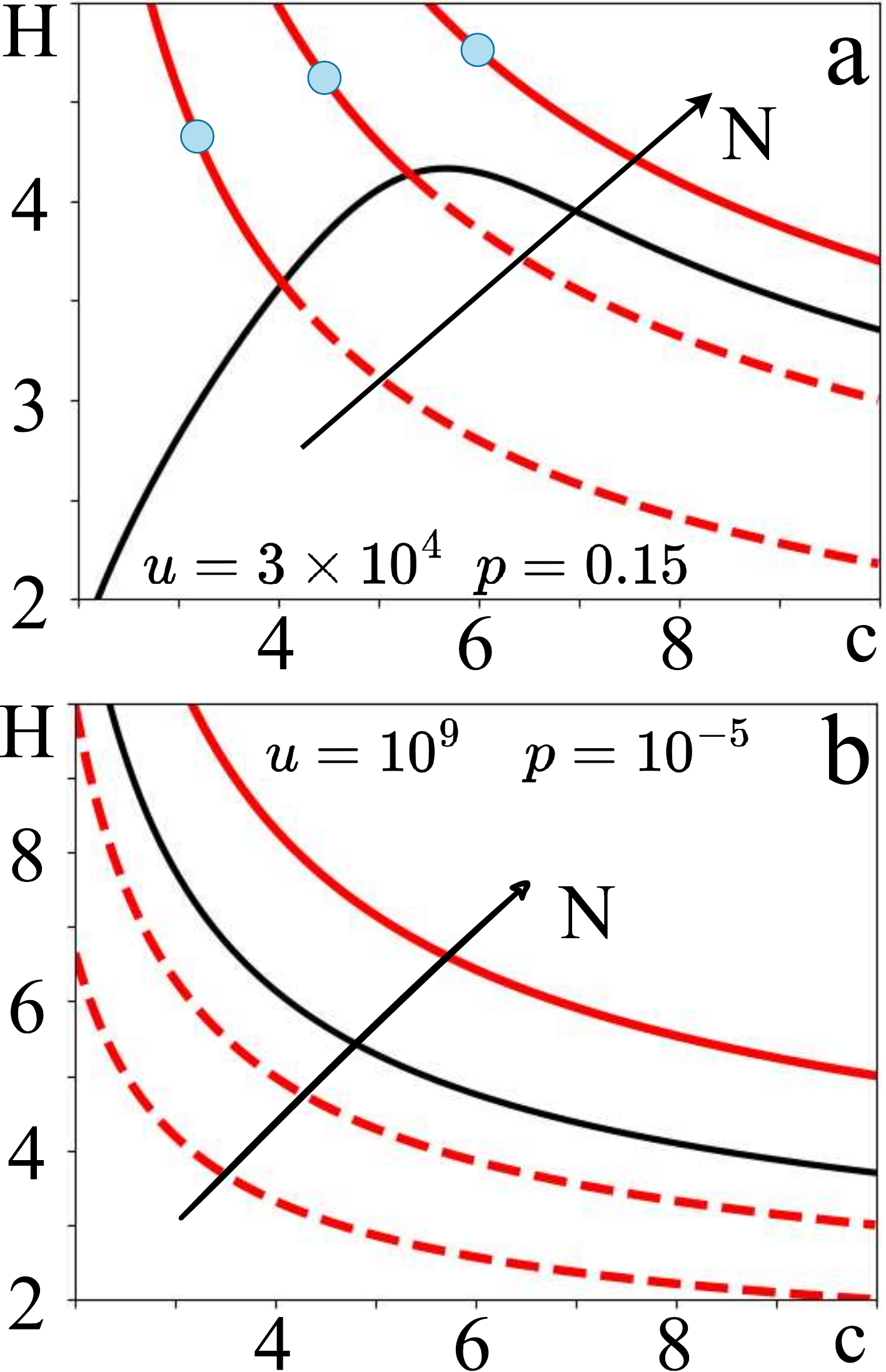}
\caption{\label{fig:N_in_c_h} The figure shows the dependence $N=c^H$ with a red line for different numbers of nodes in the graph $N$. The black line is the theoretical critical curve (\ref{approx_z_unc}) ${z}(u,p,c,H) = 1$: subfigure (a) at settings $u=3\times10^4, p=0.15$ and subfigure (b) at settings $u=10^9 p=10^{-5}$). The solid or dashed red line characterizes the dominance in this area of different search models, D-searcher and G-searcher, respectively.}
\end{figure}

\begin{figure*}[t]
\includegraphics[width=1\textwidth]{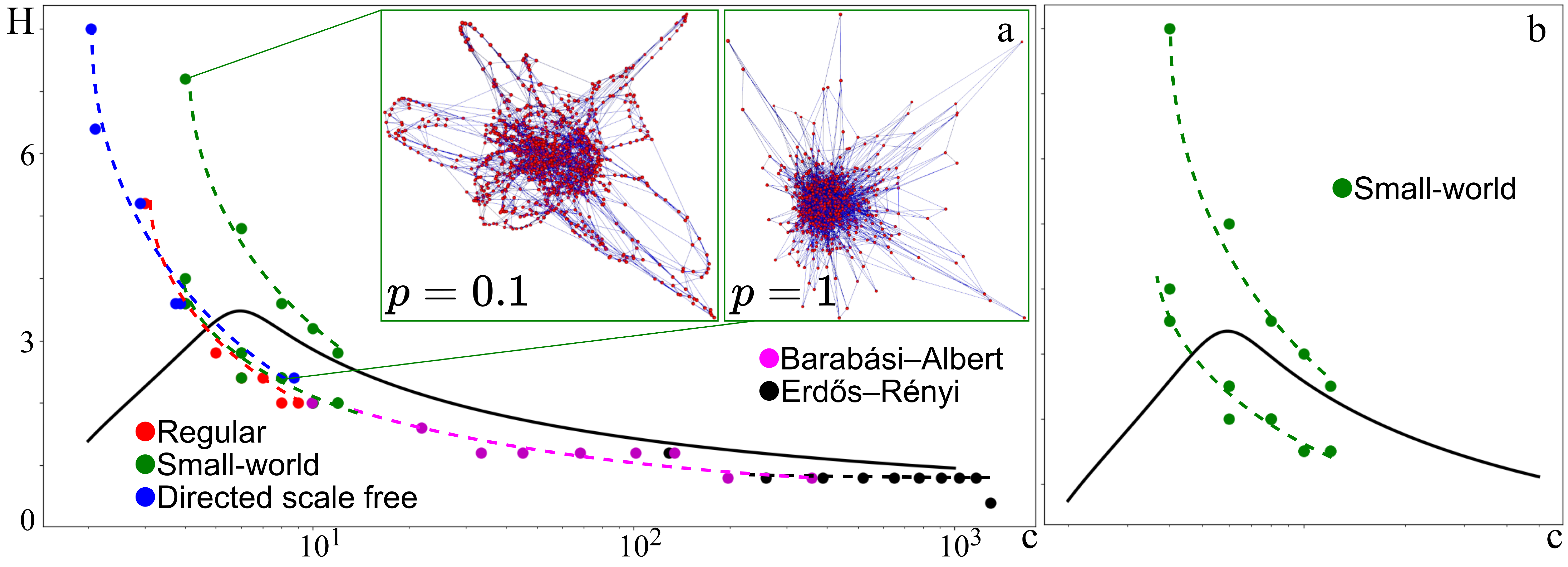}
\caption{\label{fig:graph_stability} (a) The black solid line shows the theoretical dependence \ref{approx_z_unc} with parameters $u=5 \cdot 10^4, p=0.15$ in space $(c, H)$. Each point characterizes the generated graph with the number of nodes $N=1296$. The dashed lines show the main trend of the graph properties. There are two trends for Small-world graphs characterizing different probabilities of edge rewriting $p=0.1$ left and $p=1$ right insertion. (b) A closer look at Small-world graphs. Subfigures (a) and (b) are plotted on the same axes.}
\end{figure*}

Let us briefly recall the critical line as the level curve $z=1$ on the plane $(c, H)$ which separates graphs that exhibit stable prior global search under conditions of information uncertainty from those that demonstrate the best performance with greedy random search. Our research on Erd\H{o}s-R\'enyi, random regular, and scale-free network models has shown that the critical curve behaves differently based on the parameters of the uncertainty model $(u, p)$. This curve may either monotonically decrease or transition to a bell-shaped non-monotonic line. Figure \ref{fig:N_in_c_h} illustrates two different behaviors of the curve for varying uncertainty parameters.

Our previous research focused on the stability of search algorithms for specific graphs with fixed parameters. However, we now aim to study the growth and evolution of networks over time. Specifically, we seek to determine whether it is possible to construct a growth process for a graph that exhibits desired stability properties with each added node. In other words, the graph should always be effective for either global or random search, and it should lie on one side of the critical curve during growth.
To begin this analysis, we consider a family of graphs with a fixed number of nodes $N$, represented on a plane $(c, H)$ as shown by the red lines in figure \ref{fig:N_in_c_h}. As the connectivity parameter $c$ increases, we expect the diameter of the graph to decrease to maintain a constant number of nodes. For illustration purposes, we use a simple dependency $N=c^H$ that connects the number of nodes, connectivity and graph diameter. The figure also shows the lines for different $N$, and the direction of the growth of the number of nodes in the graphs is indicated by an arrow. 
The behavior of the critical curve determines the dominant search model for the graph. In the case of a monotonic critical curve (figure \ref{fig:N_in_c_h}b), the number of nodes completely determines the stability of global search, regardless of the vertex degree in the graph. Prior to a certain threshold number of nodes $N$, global search fails, but beyond this threshold, it becomes stable. This behavior contrasts starkly with the situation when the critical curve is non-monotonic and convex upwards (figure \ref{fig:N_in_c_h}a). In this case, for all sizes $N$, there are options for graphs (solid branches with blue markers on red lines in figure \ref{fig:N_in_c_h}a) that are stable for global search in the region of small values of connectivity $c$ and large diameters $H$. We believe further development of this simple analysis may be useful for building a network growth model to maintain the stability of prior global search under incomplete information or uncertainty.

\subsection{\label{sec:flexibility_small_world_networks} Flexibility of Small-world Networks}

Previously, we drew lines corresponding to the set of different graphs with the same number of nodes only schematically. Obviously, for different graph types, they will differ, and in this section we will numerically study their behavior. Let's consider the following graph models: regular, small-world, directed scale-free,  Erd\H{o}s-R\'enyi, and Barab\'asi–Albert. We fix the number of nodes in the graph $N=1296$, and by varying different generation parameters, we obtain networks with different average connectivity and diameters. The results are shown in figure \ref{fig:graph_stability} where different graph models are marked with different colors. As we can see from the figure, random graphs of the same family  tend to lie on the same power curve. 

However, there is an exception for small-world graphs (see figure \ref{fig:graph_stability}), which are divided into several branches due to the spread in the generation parameter of the edge rewriting probability. In case the probability is $p=1$, all edges will be randomly rewritten, which reduces graphs with the small-world structure to a random regular graph. It also reduces the diameter of the graph by increasing the variance of the distribution of degrees of nodes. This is clearly observed in the left inset on figure \ref{fig:graph_stability} where the same graph transforms into different graphs depending on the parameter $p$. Therefore, by varying this parameter, different families of small-world graphs can be obtained, forming branches that demonstrate stability or instability of global search (figure \ref{fig:graph_stability}b).

\subsection{\label{sec:biological_miniaturization_associative_learning} Biological Miniaturization and Associative Learning}

Miniaturization of insects has long attracted zoologists and entomologists [\onlinecite{polilov2015small}, \onlinecite{polilov2023extremely}]. Insects are small and even very small. Among them there are those who, being a multi-cellular organism containing thousands of cells, are themselves smaller than a single-celled amoeba [\onlinecite{polilov2015small}]. So, for example, the body length of a miniature beetle is less than 400 microns; The most amazing thing is that almost all of them retain a very complex body plan, unlike other invertebrates, which lose entire organ systems during miniaturization. Of course, the miniaturization of insects affects their structure, lifestyle, and even the genome. At the same time, both general changes in body functions and individual species differences are manifested. These general changes primarily include biophysics, because when you are very small, you have a different relationship with air and water. For example, a recent study found that wing-fly microbeetles fly differently than all other animals, offsetting the loss of average speed with decreasing body size [\onlinecite{farisenkov2022novel}]. Secondly, this is the reduction of the transport and respiratory system, since at such sizes the diffusion mechanism is sufficient to transport oxygen and other substances in the body.

However, the main consequences of miniaturization are probably changes in the brain and nervous system of insects. Such miniature organisms show a disproportionate size of the nervous system and an increase in its relative volume up to 15–20 percent of body weight. The number of nerve cells is preserved, but they become very small, sometimes the nuclei disappear in the cells [\onlinecite{polilov2023extremely}]. The insect which is smaller than a unicellular organism retains hundreds of separate muscles that control the legs, wings, and mouth jaws. The nervous system cannot be very small for the accurate coordination of such a set of muscles.

Miniaturization leads to significant changes in the structure of the insect brain. Currently, one of the main tasks of entomologists is the study of memory and the possibility of associative learning in such insects [\onlinecite{fedorova2022associative}]. All this will allow you to go deeper into understanding the functioning of their brain and nervous system. At the same time, due to the size of the organisms under consideration, this is a technically complex experimental problem. Such experimental setup based on the Morris water maze design [\onlinecite{morris1981spatial}] for aversive training of various miniature insects using visual stimuli. The main idea is that the insect runs around the arena heated to a high temperature and is looking for a controlled cold spot with a comfortable temperature, while focusing on visual cues on the screen. Special software controls this - it tracks moving insects based on video cameras, changes the location of cold spots and visual images, and then processes the received data.

\begin{figure}[t]
\includegraphics[width=0.47\textwidth]{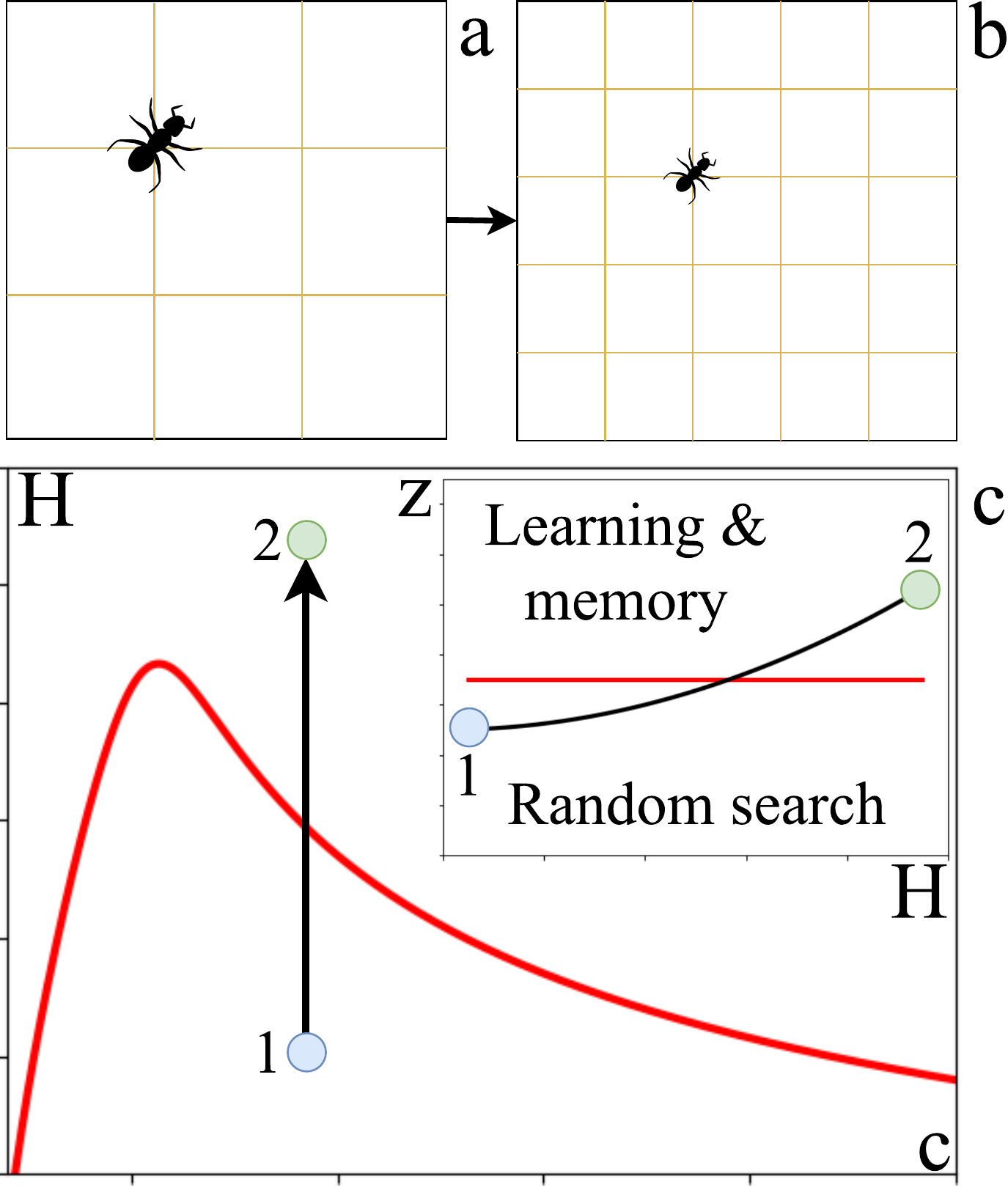}
\caption{\label{fig:sketch_scales}
(a, b) Schematic illustration the process of decreasing the body size of a searching agent. As the body size decreases, the agent can navigate through smaller spaces and access more areas of the graph. This results in the graph becoming larger without fundamentally changing its topology. (c) This figure depicts the miniaturization process on the parameter plane $(c, H)$. The arrow indicates a gradual reduction in the size of the walker agent. Symbolically, point 1 represents the graph in figure (a), while point 2 corresponds to the graph in figure (b). During this process, only the parameter $H$ changes without altering the connectivity of the graph. The critical line $z = 1$ is crossed starting from some $z_1 < 1$ and ending at $z_2 > 1$ (see insert).}
\end{figure}
 
In a recent paper [\onlinecite{fedorova2022associative}], such an experimental setup was developed and used to study the behavior of Thrips tabaci, and the ability for associative learning and memory formation was demonstrated in thrips for the first time. It turns out that miniature insects are not only able to learn, but also do it very quickly. Rats learn [\onlinecite{morris1981spatial}] in 15 to 18 attempts, crickets learn in 10, and Trichograms learn in the fourth time and start orienting themselves to visual stimuli, which is amazing.

Let us show how the acceleration of learning and cognitive abilities for global search during miniaturization can be predicted within our model. Consider an insect agent that moves in some limited area in space. For simplicity, we represent its possible movements in the form of a regular grid graph (see figure \ref{fig:sketch_scales}a). In the process of decreasing the size of the body this graph becomes larger without fundamentally changing the topology. That is, each edge in the graph reflects the possible movement of an insect in one time step and is related to the size of its body (see figure \ref{fig:sketch_scales}b).

Consider this miniaturization process on the parameter plane $(c, H)$. Recall that c represents the number of neighbors in the graph, and $H$ is the size of the graph (in terms of diameter). Thus, in figure \ref{fig:sketch_scales}c the arrow shows such a process of gradual decrease in the size of the walker agent. Symbolically, point 1 corresponds to the graph in figure \ref{fig:sketch_scales}a and point 2 corresponds to the graph in figure \ref{fig:sketch_scales}b. As you can see, in this process only $H$ changes without changing the connectivity and it crosses the critical line $z = 1$ starting from some $z_1 < 1$ and ending with $z_2 > 1$. In this figure the critical curve is plotted for some arbitrarily chosen uncertainty parameters in the graph. Naturally, the uncertainty in wildlife will be different depending on the habitat conditions of species and ecosystems. However, this is quite enough for us to demonstrate the general principle.
As a result, along the path of miniaturization, we cross the critical line (see inset in figure \ref{fig:sketch_scales}c). As described in detail above, this corresponds to a transition from a set of graphs, where stochastic G-searcher demonstrates the greatest performance, to an area where such a behavior model becomes inefficient. At the same time, the smaller the size of the agent, the more optimal and efficient the global search becomes based on the analysis of the environment. Such a strategy of searching and walking on a graph can only be implemented with strong cognitive abilities and associative memory. We emphasize that the smaller the agent, the larger $z$ and, accordingly, the need for cognitive abilities and learning speed is ever higher.
Indeed, the qualitative picture of such behavior can be explained. Microscopic beetles mainly live in decaying plant substrates and eat the bacterial and yeast film that forms there. At the same time, such insects actively fly, and are not simply carried in wind currents. But why? Because flying with the wind means moving in a random direction. And if you live inside a specific mushroom with a suitable microflora, then you need to find this mushroom in the forest. Flying with the wind, the probability of this is extremely small. It's also important that you don't have a lot of time, because outside of the substrate you will dry out because it's very small.
Thus, we believe that these coincidences are not accidental and of course require a more careful and separate study. For example, as mentioned above, the location of the red critical line in figure \ref{fig:sketch_scales}c  directly depends on the probabilistic nature of the uncertainty and may be different for different insect species in different habitats. Thus, the critical body size at which cognitive abilities begin to play a significant role in spatial movement and search may differ for different organisms in different habitat conditions. All of the above, in our strong conviction, reinforces the importance of mathematical consideration of movement strategies, taking into account the interaction with the environment of both individual agents and the collective behavior of swarms.

\section{\label{sec:conclusion}Conclusion}

The uncertainty that arises on a graph representing a particular space is closely linked to the conditions of the surrounding environment. However, determining the search algorithm that can consistently identify the shortest route on average in such uncertainty conditions is not a straightforward matter. In addition, identifying the most stable algorithm for uncertain environments and graph structures is also an intriguing question. 
We have proposed a theoretical model to investigate whether a global search algorithm with incomplete prior information can be outperformed on average by a stochastic greedy search. The model perturbs the edge weights in the graph with random variables to model the uncertainty of available information. Our results show that some graphs and uncertainty model parameters exist in which the global search algorithm fails under conditions of uncertainty, while the random greedy search performs better. We have derived a critical curve that separates graphs in which global search with incomplete information is stable from those where it fails. Interestingly, the critical curve's behavior changes from monotonic to bell-shaped, depending on the uncertainty parameters.

To test the proposed model we provide comparison with numerical simulations of two searcher models on various synthetic and real-world graphs with different structures. The synthetic graphs include random regular, n-dimensional uniform lattices, Small-world networks, Erd\H{o}s-R\'enyi, and Scale-free Barab'asi-Albert graphs, while the real graphs include a social network graph of friends based on anonymized Facebook data and a graph of the vessel system of the mouse brain. Our results show an excellent quantitative agreement between theory and numerical experiments for all graph models except those with heavy-tailed connectivity distributions and community structures. Such networks demonstrate the robustness of global search, which deviates from theoretical predictions for graphs with similar average connectivity. Therefore,  advanced model development for these graphs represents an important task and a significant interest for further research in this field.
Current research may be interesting and valuable for scientists in the field of network science, graph theory, and optimization. The study of the critical curve and its behavior for different network models and uncertainty parameters can help researchers better understand the stability of search algorithms in complex networks under conditions of incomplete information or uncertainty. Furthermore, the analysis of network growth and evolution to maintain stable search properties can be useful for designing and optimizing network architectures in various fields, such as communication networks, social networks, and biological networks.
Regarding the possible applications of this research topic, the findings could be applied in the design and optimization of search algorithms in various network-based applications. For example, in communication networks, such as the internet or wireless networks, optimizing search algorithms can improve the efficiency of information retrieval and routing. In social networks, optimizing search algorithms can enhance recommendation systems and community detection. In biological networks, understanding the stability of search algorithms can help predict and treat diseases that are influenced by network interactions, such as protein-protein interactions or gene regulatory networks.

Moreover, the study of memory and associative learning in miniature insects is currently one of the main tasks of entomologists, and it can provide insights into how the brain and nervous system of these organisms work. It was shown that acceleration of learning and cognitive abilities for global search during miniaturization can also be predicted using a model based on the possible movements of an insect agent in a regular grid graph. This can have applications in the development of efficient search and walking strategies for small robots or other devices that operate in a limited area in space.

\section{\label{sec:acknowledgement}Acknowledgement} The authors would like to express their gratitude to Mikhail Gelfand and Alexey Polilov for an impressive interview, where the latest modern results of entomologists in the study of miniature insects are clearly described in popular science language.

\nocite{*}
\bibliography{aipsamp}

\newpage
\pagebreak [5]

\begin{center}
\textbf{\large Supplemental Materials: Prior Global Search Stability on Finite Graphs with Uncertainty. May Greedy Search Win?}
\end{center}
\setcounter{equation}{0}
\setcounter{figure}{0}
\setcounter{table}{0}
\setcounter{page}{1}
\setcounter{section}{0}
\makeatletter
\renewcommand{\theequation}{S\arabic{equation}}
\renewcommand{\thefigure}{S\arabic{figure}}
\renewcommand{\bibnumfmt}[1]{[S#1]}
\renewcommand{\citenumfont}[1]{S#1}

\section{\label{MFPT_CGC} Mean First Passage Time on Coarse Grain Graph}

\begin{figure*}[t]
\includegraphics[width=0.6\textwidth]{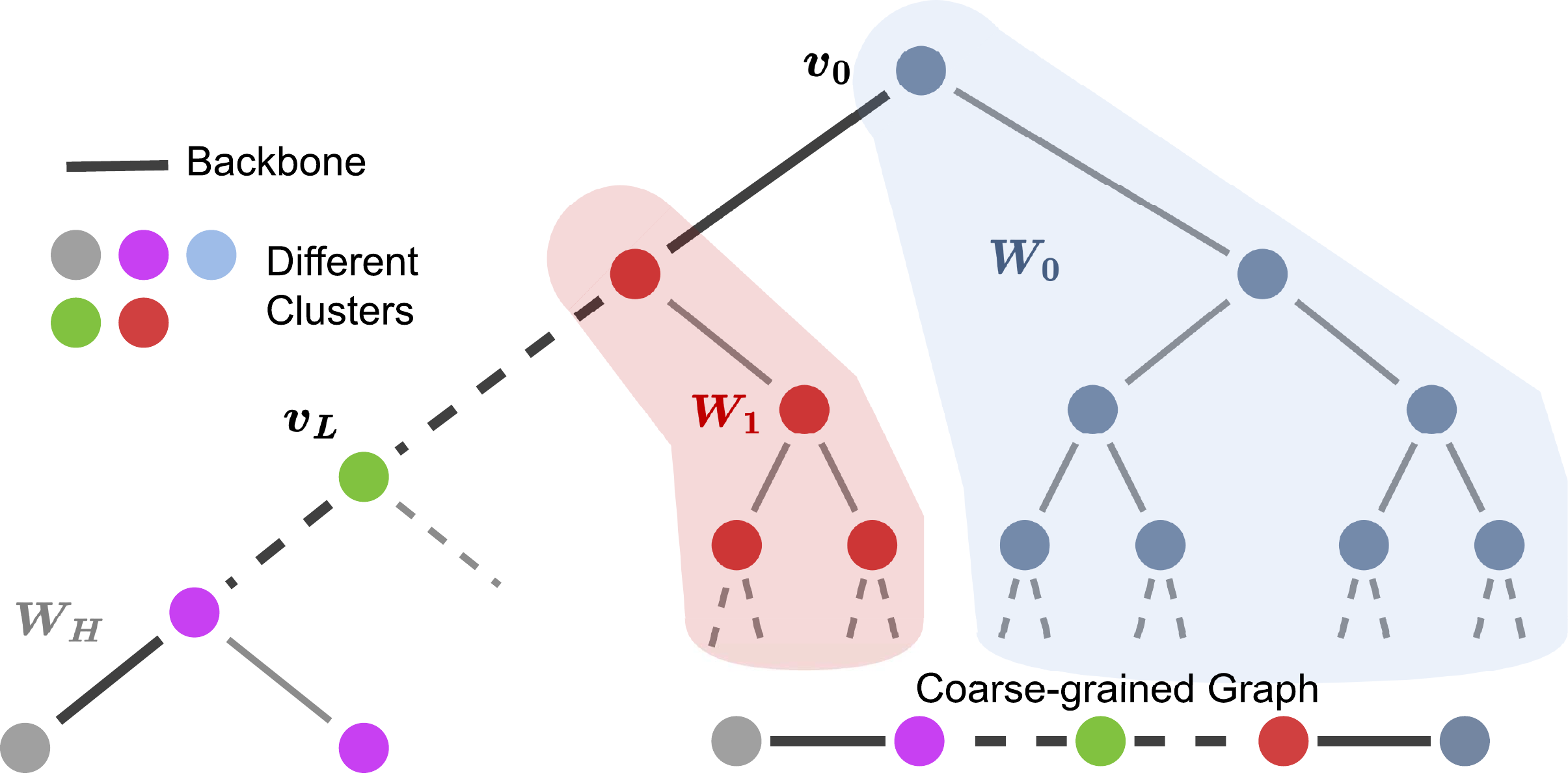}
\caption{\label{fig:c-tree} An example of a tree graph represented as a coarse grained graph. Nodes marked with the same color belong to the same cluster. $W_0$ and $W_1$ - weights of 0 and 1 clusters, respectively. The dotted line represents the rest of the graph.}
\end{figure*}

Below, we will present the proof of equation (\ref{mfpt_tree_exact}). For the beginning, let's introduce the basic concepts. The graph is undirected: $e_{i j} = e_{j i}$. The graph can be represented as a coarse-grained graph by grouping the nodes of the original network into clusters. Moreover, each tree can be reduced to a coarse grained graph (see fig. \ref{fig:c-tree}). $C$ is the transition matrix from the original graph to coarse-grained, which has size $N \times M$, where $M$ is count of clusters. And $C_{i,I} = 1$ if the $i$ node is in the $I$ cluster, and $C_{i,I} = 0$ otherwise. Here and below, we denote the indices related to the cluster number using uppercase characters. 

Also, we define the vector of transition probabilities characterizing the equilibrium state $\boldsymbol{\pi}$ for the original network: $\pi_i = \frac{k_i}{\sum_j k_j}$, where ${\pi}_{i} p_{ij} = {\pi}_j p_{ji}$ and for the coarse-grained graph:
\begin{equation}
\Pi_K = \sum_{j=0}^{|K|-1} \pi_{v_{K_j}}
\end{equation}
where ${\Pi}^T = {\pi}^T C$. This allows us to denote a transition probability matrix for a coarse-grained graph:
\begin{equation}
P_{IJ} = \frac{1}{\Pi_I}\sum_{ij} C_{iI} \pi_i p_{ij} C_{jJ}
\end{equation}
Before moving on, let's determine the weight of the tree $\boldsymbol{t}$:
\begin{equation}
     \omega (\boldsymbol{t}) = \prod_{(ij) \in \boldsymbol{t}} p_{ij}
\end{equation}

In the paper [\onlinecite{forster2022exact}], [\onlinecite{chebotarev2007graph}], [\onlinecite{pitman2018tree}], proposes a proof based on a combinatorial approach to calculating the MFPT, which consists in finding all spanning trees and forests of two trees in a graph. The MFPT from node $i$ to $j$ is found
within this approach
\begin{equation}
     m_{ij} = \frac{s_{ij}}{s_j}
     \label{mfpt_s}
\end{equation}
where $s_j$ the sum of all possible spanning trees rooted at the node $j$:
\begin{equation}
     s_{j} = \sum_{\boldsymbol{t} \to j} \omega(\boldsymbol{t})
\end{equation}
and $s_{ij}$ is the sum of the weights of all two-tree forests $(\boldsymbol{t}, \boldsymbol{s})$, such that $\boldsymbol{t}$ has root $j$ and $\boldsymbol{s}$
contains $i$
\begin{equation}
     s_{ij} = \sum_{\boldsymbol{t} \to j; i \in \boldsymbol{s}} \omega (\boldsymbol{t}) \omega (\boldsymbol{s})
     \label{two_tree_s}
\end{equation}
Using the introduced quantities, one can express the stationary probabilities of a random walker

\begin{equation}
    \pi_j = \frac{s_j}{\sum^n_{k=1} s_k}
\end{equation}
For expression $m_{v_0 v_H}$, it is necessary to find all spanning trees that will participate in the sum. But each spanning tree will contain edges from $v_0$ to $v_H$, since these are the edges of the backbone in the necklace structure. So it contains a common factor, which can be denoted as follows:

\begin{eqnarray}
\omega_{I J} = 
 \begin{cases}
   \prod_{n=I}^{J-1} p_{v_n v_{n+1}} & I < J\\
   \prod_{n=J}^{I-1} p_{v_{n+1} v_{n}} & I > J\\
   1 & I=J
 \end{cases}
\end{eqnarray}
Denote the sum of all spanning trees from subgraph $I$ rooted at node $v_I$ as $\omega(I \to v_I)$. Since the backbone and all subgraphs with roots at its nodes make up all the vertices, then $s_{v_H}$ can be written as:
\begin{equation}
    s_{v_H} = \omega_{0,H} \prod_{I=0}^H  \omega(I \to v_I)
\end{equation}
For using equation (\ref{two_tree_s}), we divide the graph into two trees by removing the edge of the backbone $(v_{I-1}, v_I)$, where $I > 1$. All other graph edges must be included. The weight of the subgraph including the $H$ node after removing the edge: $\omega_{IH} \prod_{J=I}^H \omega(J \to v_J)$. For another tree that does not include node $v_H$, it is necessary to find all remaining trees with all possible roots of $v_{K_m}$ in the corresponding subtrees $K$. This allows us to write the numerator of the equation (\ref{mfpt_s}) in the following form:

\begin{widetext}
\begin{eqnarray}
s_{v_0 v_H} & = \sum_{I=1}^H \left[ \omega_{IH} \prod_{J=I}^H \omega(J \to v_J) \times  \sum_{K=0}^{I-1} \left( \sum_{m=0}^{|K|-1} \omega(K \to v_{K_m}) \omega_{0 K} \omega_{I-1, K} \prod_{J=0, J \neq K}^{I-1} \omega(J \to v_J) \right) \right] = \nonumber \\ 
& = \sum_{I=1}^H \sum_{K=0}^{I-1} \left[ \omega_{IH} \omega_{0K} \omega_{I-1,K} \frac{\prod_{J=0}^H \omega(J \to v_J)}{\omega(K \to v_K)} \sum_{m=0}^{|K| - 1} \omega(K \to v_{K_m}) \right]
\end{eqnarray}
\end{widetext}
Without loss of generality, we can use the obtained expressions to search for MFPT between any nodes on the backbone. Furthermore, searching for the MFPF between the first node of backbone and any node $v_i$ is done by selecting the backbone so that it includes the target node of interest. This means that the node $v_i$ will be the root for all spanning trees in the corresponding  subgraph and will persist as a cluster in the coarse grained graph. So the equation (\ref{mfpt_s}) for nodes $v_0$ and $v_L$ looks like this:

\begin{eqnarray}
m_{v_0 v_L} & =& \sum_{I=1}^L \sum_{K=0}^{I-1}  \frac{ \omega_{IL} \omega_{0K} \omega_{I-1,K} }{\omega_{0L} \prod_{J=0}^L \omega(J \to v_J)} \nonumber \\
& \times&  \frac{\prod_{J=0}^L \omega(J \to v_J)}{\omega(K \to v_K)} \sum_{m=0}^{|K|-1}\omega(K \to v_{K_m}) \nonumber\\ & =&  \sum_{I=1}^L \sum_{K=0}^{I-1} \frac{\omega_{I-1,K}}{\omega_{KI}} \frac{\Pi_K}{\pi_{v_K}}
\end{eqnarray}
where $v_L$ - is the tree node at depth $L$, which corresponds to the distance between these nodes.

Using transition probability on a directed graph, $p_{ij} = \frac{e_{ij}}{k_i}$, let's transform  the first factor of this sum:
\begin{eqnarray}
{\frac{\omega_{I-1,K}}{\omega_{KI}}} & = & \frac{\prod_{n=k}^{I-2} p_{v_{n+1} v_n}}{\prod_{n=k}^{I-1} p_{v_{n} v_{n+1}}} \nonumber\\
& = &\frac{ p_{v_{k+1} v_{k}} \cdot p_{v_{k+2} v_{k+1}} \cdot \ldots \cdot p_{v_{I-2} v_{I-3}} \cdot p_{v_{I-1} v_{I-2}} }{p_{v_{k} v_{k+1}} \cdot p_{v_{k+1} v_{k+2}} \cdot \ldots \cdot p_{v_{I-2} v_{I-1}} \cdot p_{v_{I-1} v_{I}}} \nonumber\\
& = &\frac{p_{v_{I} v_{I-1}}}{p_{v_{I} v_{I-1}}} \cdot \frac{\prod_{n=K}^{I-2}}{\prod_{n=K}^{I-1}} = \frac{1}{p_{v_{I} v_{I-1}}} \prod_{n=K}^{I-1} \frac{ p_{v_{n+1} v_n} }{ p_{v_{n} v_{n+1}} } \nonumber\\
& = & \frac{k_{v_I}}{e_{I I-1}} \prod_{n=K}^{I-1} \frac{e_{v_{n+1} v_{n}}}{k_{v_{n+1}}} 
\frac{k_{v_{n}}}{e_{v_{n} v_{n+1}}} \nonumber \\
& = & \frac{k_{v_I}}{e_{I I-1}} \frac{ k_{v_K} \cdot k_{v_{K+1}} \cdot \ldots \cdot k_{v_{I-1}} }{ k_{v_{K+1}} \cdot \ldots \cdot k_{v_{I-1}} \cdot k_{v_{I}} } = k_{v_K}
\end{eqnarray}
Also, for the second factor in this sum:
\begin{eqnarray}
m_{v_0 v_L} & = &  \sum_{I=1}^L \sum_{K=0}^{I-1} k_{v_K} \frac{\sum_{i=0}^{|K|-1} \pi_{v_{K_i}}}{\pi_{v_K}} \nonumber\\
& = & \sum_{I=1}^L \sum_{K=0}^{I-1} k_{v_K} \sum_{i=0}^{|K|-1} \frac{k_{v_{K_i}}}{\boldsymbol{W}} \frac{\boldsymbol{W}}{k_{v_K}} \nonumber\\
& = & \sum_{I=1}^L \sum_{K=0}^{I-1} \sum_{i=0}^{|K|-1} k_{v_{K_i}}
\label{mfpt_3sum}
\end{eqnarray}
The next step is calculate first internal sum of equation (\ref{mfpt_3sum}). Let's denote this amount as $W_K = \sum_{i=0}^{|K|-1} k_{v_{K_i}}$. This is the sum of the degrees of the vertices within the cluster. For any graph it is true that $\boldsymbol{W} = \sum_{i} k_i = 2|E|$. This statement can be supplemented for $c$-trees like this $\boldsymbol{W}=2(N-1)$. 

The sum of all nodes degrees for a cluster at the root of $c$-ary tree is the sum of all nodes degrees for the whole tree minus the nodes degrees in one child subtree.	 Conversely, the sum of the degrees of the nodes in the subtree at height $K$ is the sum of the degrees in the tree of height $H-K$ and the one corresponding to the edge leading to the root of the tree (see fig. \ref{fig:c-tree}).

First calculate the sum of the vertex degrees for $c$-ary tree with the height $H$  using the well-known relationship [\onlinecite{CaryTree}] linking $c$-ary tree depth to the number of nodes $N_c(H) = \frac{c^{H+1} - 1}{c-1}$, 
\begin{equation}
    \boldsymbol{W_c^H}  =  2(N_c(H)-1) = \frac{2c(c^H-1)}{c-1}    
\end{equation}

This makes it possible to calculate $W_0$ as described in the paragraph above:
\begin{equation}
    W_0 = \boldsymbol{W_c^H} - (\boldsymbol{W_c^{H-1} + 1)} 
     =  2c^H - 1
\end{equation}
For the remaining clusters the situation is similar, but each cluster has an additional link edge to the previous one: 
\begin{eqnarray}
    W_K & = & \frac{2c(c^{H-K} - 1)}{c-1}-\frac{2c(c^{H-K-1} - 1)}{c-1} + 1 \nonumber\\ 
    & = & 2c^{H-K}
\end{eqnarray}
The found sums of vertex degrees across clusters allow us to return to the basic equation (\ref{mfpt_3sum}) and continue calculating MFPT:

\begin{eqnarray}
    m_{v_0 v_L} & = & \sum_{I=1}^{L} \sum_{K=0}^{I-1} W_K = L W_0 + \sum_{I=1}^{L} \sum_{K=1}^{I-1} W_K \nonumber \\
    & = & L \left( 2c^H - 1\right) + \sum_{I=1}^L \sum_{K=1}^{I-1} 2 c^H \frac{1}{c^K}
\end{eqnarray}
Using simple algebra and revealing all the sums, we obtain the following equation on $m_{v_0 v_L}$
\begin{eqnarray}
    m_{v_0 v_L} & = &  L \left( 2c^H - 1\right) + 2 c^H \sum_{I=1}^L \frac{c^{-I} \left( c^I - c \right) }{c-1} \nonumber\\
    & = & L \left( \frac{2c^{H+1}}{c-1} - 1\right) - \frac{2 c^{H+1} (1 - c^{-L}) }{(c-1)^2}
    \label{to_ending_m}
\end{eqnarray}

The resulting expression is an exact formula for calculating the MFPT on the tree. Moreover, this expression can be simplified for large graphs ($c \gg 1, H \gg 1$). 
Use the first condition ($c \gg 1$) to simplify the numerators in the main expression (\ref{to_ending_m})
\begin{equation}
m_{v_0 v_L}   \approx   \frac{2Lc^{H+1}}{c-1} - \frac{2c^{H+1}}{(c-1)^2}
\end{equation}
Now using the second condition ($H \gg 1$) simplify the expression and get rid of the denominators
\begin{equation}
m_{v_0 v_L}   \approx  2Lc^{H} - 2c^{H-1}
 \approx  2Lc^H  = \tilde m_{v_0 v_L}
\end{equation}

thus we obtain the final expression.

\section{\label{sim_alg} Simulation algorithms}

The basic cycles for obtaining numerical values of $z(c,H)$ and $z(u,p)$ via simulations are demonstrated in A\ref{alg:z_cd} and A\ref{alg:z_up} respectively.  The first cycles are varied by the corresponding parameters. The last one is necessary for averaging the results. For the purposes of this work, all results are averaged over 1000 runs. Based on the fixed parameters, a graph is generated and a basic simulation is run, which is described in A\ref{alg:simulation}.

\begin{algorithm}[H]
\SetAlgoLined
\DontPrintSemicolon
\label{alg:z_cd}
\caption{Calculate $z(c,H)$}
\KwData{$c_{max}, H_{max}, u, p, N - \text{number of runs}$}
\For{$c\gets2$ \KwTo $c_{max}$}{
    \For{$H\gets2$ \KwTo $H_{max}$}{
        $z_{tmp} \gets 0$\;
        \For{$i\gets0$ \KwTo $N$}{ 
            $G \gets GraphGenerator(c,H)$\;
            $z_{tmp} \gets z_{tmp} + Simulation(G,u,p)$\;
        }
        $z[c,H] \gets \frac{z_{tmp}}{N}$\;
    }
}
\end{algorithm}

As part of the basic simulation, nodes are randomly selected on the graph. Between the selected nodes the minimal unweighted route is searched $Parh_{D}$. Then moving along this route repeatedly for each edge its weight is generated and added to the total weight of the route $\mathbb{W}_{s,t}^D$.

\begin{algorithm}[H]
\SetAlgoLined
\DontPrintSemicolon
\label{alg:z_up}
\caption{Calculate $z(u,p)$}
\KwData{$c, H, d_{max}, p_{max}, p_{step}, N - \text{number of runs} $}
$G \gets GraphGenerator(c,H)$\;
\For{$d\gets2$ \KwTo $d_{max}$}{
    \For{$p\gets0$ \KwTo $p_{max} \textbf{ step } p_{step}$}{
        $z_{tmp} \gets 0$\;
        \For{$i\gets0$ \KwTo $N$}{  
            $u\gets 10^d$\;
            $z_{tmp}\gets z_{tmp} + Simulation(G,u,p)$\;
        }
        $z[u,p] \gets \frac{z_{tmp}}{N}$\;
    }
}
\end{algorithm}

Next, the route and weight for the G-searcher are calculated. The source is marked as the current vertex $n_{cur}$. Next, its weight is sampled for each incident edge. The edge with the smallest weight is selected from all incident edges. If there are several such edges, then an edge is selected randomly from them. Next, the walker moves to the next node along the selected edge. The weight of the edge is added to the main weight of the route $\mathbb{W}_{s,t}^G$. This happens until the walker reaches the target vertex.

\begin{algorithm}[H]
\SetAlgoLined
\DontPrintSemicolon
\label{alg:simulation}
\caption{Simulation$(G,u,p)$}
\KwData{$G$ - graph, $u$ - an extra weight, $p$ - probability of realization}
\KwResult{$z$}
$source \gets \text{pick random node from G}$\;
$target \gets \text{pick random node from G}$\;
$Path_D \gets \text{run Dijkstra alg. between $source$ and $target$}$\;
$\mathbb{W}^{D}_{s,t} \gets 0$\;
\For{$i\gets0$ \KwTo $length(Path_D) - 1$}{    
    \If{$\text{random} \xi < p$}{
        $\mathbb{W}^{D}_{s,t} += 1 + u$\;
    }
}
$n_{cur} \gets source$\;
$\mathbb{W}^{G}_{s,t} \gets 0$ \;
\While{$n_{cur} != target$}{
    $n_{cur} \gets \text{transition to the vertex to the edge with}$\;
    $\text{the smallest weight $W_{min}$}$\;
    $\mathbb{W}^{G}_{s,t} += 1 + W_{min}$\;
}
$\textbf{return} \frac{\mathbb{W}^{G}_{s,t}}{\mathbb{W}^{D}_{s,t}}$\;

\end{algorithm}

\end{document}